\documentclass{ieeeaccess}
\usepackage{cite}
\usepackage{amsmath,amssymb,amsfonts}
\usepackage{algorithmic}
\usepackage{graphicx}
\usepackage{textcomp}

\usepackage[table,xcdraw]{xcolor}
\usepackage{xspace}
\usepackage{spverbatim}
\usepackage{booktabs}
\usepackage{rotating}
\usepackage{multirow}
\usepackage{tablefootnote}
\usepackage[nottoc]{tocbibind}
\usepackage{threeparttable}
\usepackage{multicol}
\usepackage{listings}
\usepackage{hyperref}

\usepackage{bm}
\makeatletter
\AtBeginDocument{\DeclareMathVersion{bold}
\SetSymbolFont{operators}{bold}{T1}{times}{b}{n}
\SetSymbolFont{NewLetters}{bold}{T1}{times}{b}{it}
\SetMathAlphabet{\mathrm}{bold}{T1}{times}{b}{n}
\SetMathAlphabet{\mathit}{bold}{T1}{times}{b}{it}
\SetMathAlphabet{\mathbf}{bold}{T1}{times}{b}{n}
\SetMathAlphabet{\mathtt}{bold}{OT1}{pcr}{b}{n}
\SetSymbolFont{symbols}{bold}{OMS}{cmsy}{b}{n}
\renewcommand\boldmath{\@nomath\boldmath\mathversion{bold}}}
\makeatother

\def\BibTeX{{\rm B\kern-.05em{\sc i\kern-.025em b}\kern-.08em
    T\kern-.1667em\lower.7ex\hbox{E}\kern-.125emX}}

\newcommand{\PPA}{AI-driven PPA\xspace}

\newcommand{\PPAs}{AI-driven PPAs\xspace}

\begin{document}
\history{Date of publication xxxx 00, 0000, date of current version xxxx 00, 0000.}
\doi{10.1109/ACCESS.2024.0429000}

\title{AI-driven Personalized Privacy Assistants: a Systematic Literature Review}
\author{\uppercase{Victor Morel}\authorrefmark{1}, 
\uppercase{Leonardo Horn Iwaya}\authorrefmark{2}, and \uppercase{Simone Fischer-Hübner}\authorrefmark{1,2}}

\address[1]{Chalmers University of Technology and University of Gothenburg, Gothenburg, Sweden}
\address[2]{Department of Mathematics and Computer Science, Faculty of Health, Science, and Technology, Karlstad University, Karlstad, Sweden}
\tfootnote{This work was partially supported by the Wallenberg AI, Autonomous Systems and Software Program (WASP) funded by the Knut and Alice Wallenberg Foundation.
The work of Iwaya L.H. is partly supported in part by the
Knowledge Foundation of Sweden (KKS), Region V\"{a}rmland (Grant: RUN/230445) and the European Regional Development Fund (ERDF) (Grant: 20365177) in connection to the DHINO 2 project, and Vinnova (Grant: 2018-03025) via the DigitalWell Arena project.}

\markboth
{Morel \headeretal: AI-driven Personalized Privacy Assistants: a Systematic Literature Review}
{Morel \headeretal: AI-driven Personalized Privacy Assistants: a Systematic Literature Review}

\corresp{Corresponding author: Victor Morel (e-mail: morelv@chalmers.se).}

\begin{abstract}
In recent years, several personalized assistants based on AI have been researched and developed to help users make privacy-related decisions.
These AI-driven Personalized Privacy Assistants (\PPAs) can provide significant benefits for users, who might otherwise struggle with making decisions about their personal data in online environments that often overload them with different privacy decision requests.
So far, no studies have systematically investigated the emerging topic of \PPAs, classifying their underlying technologies, architecture and features, including decision types or the accuracy of their decisions.
To fill this gap, we present a Systematic Literature Review (SLR) to map the existing solutions found in the scientific literature, which allows reasoning about existing approaches and open challenges for this research field.
We screened several hundred unique research papers over the recent years (2013-2025), constructing a classification from 41 included papers.
As a result, this SLR reviews several aspects of existing research on \PPAs in terms of types of publications, contributions, methodological quality, and other quantitative insights.
Furthermore, we provide a comprehensive classification for \PPAs, delving into their architectural choices, system contexts, types of AI used, data sources, types of decisions, and control over decisions, among other facets.
Based on our SLR, we further underline the research gaps and challenges and formulate recommendations for the design and development of \PPAs as well as avenues for future research.
\end{abstract}

\begin{keywords}
artificial intelligence, data protection, machine learning, privacy, privacy assistant, systematic review
\end{keywords}

\titlepgskip=-21pt

\maketitle

\section{Introduction}
\label{sec:introduction}
% \todo[inline]{Strengthen the motivation → @victor (substantial amount of work, novelty}
%anchor
As the world becomes increasingly digitalized, people are faced with a growing number of requests for decisions related to their online privacy.
%We surround ourselves 
Nowadays, individuals are using several apps every day, visiting different websites, and the number of smart gadgets and Internet of Things (IoT) devices they use continues to grow~\cite{noauthor_state_2024}.
Furthermore, 
%for enforcing the individuals' rights to informational self-determination and 
to comply with privacy laws such as the General Data Protection Regulation (GDPR)~\cite{european_parliament_general_2016}, software systems %regularly require 
frequently demand from us to make privacy-related decisions regarding our personal data: \textit{Do you grant this permission? Do you want to accept the cookies? Should this sensor be left on when you host friends?}
Consequently, the cognitive burden increases, leaving users in disarray, tired, and unable to decide in their best interests~\cite{choi_role_2018}.

%motivation
During the last decade, researchers have been building privacy assistants to alleviate this burden and support users in their decisions.
%(see 
One of the first research work in that field has resulted in the patent on a Personalized Privacy Assistants (PPAs), registered in 2023 in the United States by Sadeh et al.~\cite{sadeh2021personalized}).
With the progress made in Artificial Intelligence (AI), it comes at no surprise that many of 
the privacy assistants proposed or developed in recent years leverage AI technology, notably to enable better personalized support.

This personalization can enhance the quality of decision support, adapted to the individuals' needs, preferences and current context.
%brings new insights for the development of PPAs, enabling bespoke and contextual help.
However, the extent to which AI drives these \PPAs, their efficiency, privacy-friendliness, functionality, and how legal requirements are eventually addressed remains unclear.
In fact, to the best of our knowledge, there have been no surveys or systematic reviews on the topic of \PPAs, despite the substantial amount of work published on the topic in the recent years.
This lack of systematization of knowledge makes it more difficult for other researchers and developers to reason about the opportunities that current \PPAs may offer.
It also makes it more difficult to identify limitations and existing gaps to be addressed, as well as open research challenges that remain for the future.

%objective
%Addressing this research gap
For this reason, we present a Systematic Literature Review (SLR) of the body of knowledge to provide a common vocabulary and better compare, categorize and analyze the different \PPA solutions.
In doing so, we aim to draw insights and lessons for future assistants and to formulate better recommendations for research, design, and development of \PPAs.
Therefore, this SLR addresses the following Research Questions (RQs):
\begin{itemize}
    \item \textbf{RQ1:} \textit{What is the current state of the literature on \PPAs for automated support of end-users privacy decisions in IT systems?}
    \item \textbf{RQ2:} \textit{What are the key attributes and properties of the proposed \PPAs in the literature?}
\end{itemize}

Here, we consider \textbf{agents} and \textbf{assistants} in a broad sense (any logical entity able to support users, including unimplemented theoretical models, see our selection criteria in Table~\ref{tab:selection-criteria}); \textbf{AI} in a generic sense as well (see Section~\ref{subsec:AI}); and, \textbf{privacy decisions} as the individual's decisions regarding their personal information management (see Section~\ref{subsec:privacy_decisions}).

%scope
To address our RQs, we performed an SLR on research papers providing technical solutions, published between 2013 and 2025 in peer-reviewed venues, including a snowballing process until early February 2025.
We screened several hundred papers from IEEE, ACM, Scopus, and Web of Science, resulting in 41 selected papers after several rounds of snowballing.
We extensively read and analyzed all included papers, and the information extracted forms the basis of our work.

%contributions
Our SLR results in the following contributions:
\begin{itemize}
    \item \textbf{A Classification for \PPAs} -- We propose the first classification for \PPAs, providing a common vocabulary for designers of such systems.
    \item \textbf{Data Charting \& Quantification} -- We charted and quantified several aspects of \PPAs based on the aforementioned classification. 
    \item \textbf{Research Gaps \& Challenges} -- We underline the current gaps in the state of the art and highlight challenges for designing \PPAs based on our data.
    \item \textbf{Recommendations \& Research Avenues} -- We formulate recommendations for improving \PPAs, and propose several avenues for future research.
\end{itemize}

In the following sections, we present the background and related work in Section \ref{sec:background}. The study's methodology is detailed in Section \ref{sec:methodology}. The results and classification are organized and presented in Sections \ref{sec:results} and \ref{sec:classification}. Based on the findings, we present our discussion of research gaps and future work in Section \ref{sec:discussion}. Lastly, Section \ref{sec:conclusion} concludes our work.

\section{Background}
\label{sec:background}
As a background, this section first provides an overview of different types of privacy decisions for which individuals could receive support from \PPAs. 
Then, it summarizes legal requirements for transparency that \PPAs should meet, and refers to a classification scheme of explainable AI that we are using for our classification.

%presents AI and Machine Learning (ML) technologies that can provide the technical foundations for \PPAs.
When discussing legal requirements, we will primarily refer in the section and for the rest of this paper to the European Legal Framework, including the GDPR and the AI Act~\cite{eu_ai_act_2024}, since the study was conducted in Europe with the support of a European funding foundation. 
Moreover, the GDPR has been regarded as the ``gold standard'' for data protection with a territorial scope that goes beyond Europe, and is therefore also used as a point of reference.

\subsection{Privacy Decisions}
\label{subsec:privacy_decisions}
Among the most notable definitions, Westin~\cite{westin_privacy_1968} has defined privacy as the right to informational self-determination, meaning that individuals should have the \textit{right to decide} for themselves when, how, and what information about them is communicated to others.
As mentioned, in the EU, the GDPR emphasizes that individuals should have control of their personal data (Recital 7), and thus should be empowered to make decisions about their data as one prerequisite for exercising such control.
Delving deeper into this notion of \textit{privacy decisions}, we further elaborate on this concept in the following subsections.

\subsubsection{Individual Privacy Decisions Regulated by Privacy Laws}
Some privacy decisions individuals can make to exercise control over their data is regulated under the GDPR and other privacy laws. 
These decisions notably include, but are not limited to, the \textit{decisions to grant or to withdraw consent} to data collection and processing.
Art. 4 (11) of the GDPR defines ‘consent’ of data subjects as any freely given, specific, informed, and unambiguous indication of the data subject's wishes by which they, by a statement or by a clear affirmative action, signifies agreement to the processing of personal data relating to them.

Moreover, the GDPR and most other privacy laws regulate further \textit{decisions to exercise data subject rights} granted by the respective laws. 
For instance, according to Art. 15-22 GDPR, data subjects have the rights to access data, request rectification or deletion of data, export data, and object to direct marketing and profiling.
Data subjects can also object in cases where the legal ground for the processing is public interest or legitimate interest, or exercise their right not to be subject to automated decision-making.

\subsubsection{Further Types of Privacy Decisions}
\label{subsec:further-decisions}
Further types of privacy decisions concerning users' choices regarding the use of their data by others, which are not directly mentioned or regulated by the GDPR, include \textit{decisions of individuals to publish or share data on their own initiative}, e.g., in social networks. In these cases, data sharing has typically not been formally triggered by a consent request to allow data sharing with another party.
%.Moreover, privacy decisions for controlling the disclosure or conditions for the processing one's personal data include 

Moreover, privacy decisions encompass \textit{privacy permission} settings (or access control rights), which grant others certain rights for using their data and are, for instance, typically used for permission systems of mobile phone operating systems, such as Android or iOS.
Setting privacy permissions on mobile operating systems often requires consent at installation or during runtime. 
However, instead of consent, other legal grounds -- such as a contract (Art. 6 (1)(b) GDPR) --, can be used, e.g., for a banking app to forward account information when transferring money~\cite{Art29WP13}.
Let us also note the peculiar case of Global Privacy Control (GPC), a unary signal that permits or prohibits third-party tracking on the browser~\cite{human_data_2022}.
Due to its enforceability under the California Consumer Privacy Act (CCPA)~\cite{CCPA_2018}, it is regulated by a privacy law but is technically more akin to a privacy permission.

Additionally, some privacy-enhancing technologies and protocols allow users to decide and set \textit{privacy preferences}, which are simply indications of the users' privacy wishes of how their data should be used without actually granting any rights to others, and thus without legal mandate. 
Privacy preferences have, for instance, been used earlier by the Platform for Privacy Preferences (P3P)~\cite{cranor_platform_2002} or Do Not Track (DNT)~\cite{DNT_W3C_2019}, as an example for signals that can be set manually in browser settings for allowing users to specify their privacy choices.

\subsection{Requirements for Transparency.}
\label{legal_background}
Transparency of data processing is an important prerequisite for users for making well-informed decisions, and should therefore be provided by any privacy assistants that should support users in decision-making.
In cases where the data controllers of the \PPAs are not the data subjects themselves, the controllers should provide the data subjects with privacy policy information \textit{ex-ante} at the time when data is obtained from them according to Art. 13 GDPR, and \textit{ex-post} through the right to access granted in Art. 15 GDPR.
This should particularly include information about purposes of processing, data categories concerned, but also information about the logic involved and significance, and envisioned consequences of automated decision-making and profiling performed by the \PPAs.

The EU AI Act also includes obligations for transparency for the producers and deployers of limited-risk and high-risk AI systems (Art. 50). While the providers of limited-risk AI systems have to mainly ensure that humans are informed that AI systems are used, high-risk AI systems require that further clear, comprehensible and adequate information is given to the deployer (Art. 13), traceability of results via logging (Art. 12) and appropriate human oversight (Art. 14). However, \PPAs are typically not in the high-risk category, since they are used for users' own personal privacy management, which should typically not interfere with the fundamental rights of others. Exceptions could, however, be \PPAs that are, for example, used for setting permissions for safety-critical applications impacting the safety of the users or others.

Also, Ethics Guidelines for Trustworthy AI, promoted by the EU Commission,\footnote{\url{https://digital-strategy.ec.europa.eu/en/library/ethics-guidelines-trustworthy-ai}} emphasize the requirement for transparency and explainability for AI systems to be deemed trustworthy.

\subsection{AI for Decision-making}
\label{subsec:AI}
AI is a generic term for various strategies and techniques enabling computers and machines to simulate human intelligence and problem-solving capabilities~\cite{russell_artificial_2016}.
Machine learning (ML) is a field of AI (we subsume the former under the latter in the rest of the document) that develops and studies statistical algorithms and models, draws inferences from patterns in data, and learns and adapts without following explicit instructions.
AI-powered tools can particularly lighten the user's cognitive load and thereby improve their decision-making, e.g., by decision support, augmentation, or automation. 

While there are different ways to categorize AI systems, we refer in the present work to the survey paper on eXplainable AI (XAI) by Arrieta et al.~\cite{arrieta_explainable_2019}.
They distinguish between transparent models and those requiring post-hoc explainability (called \textit{non-inherently transparent}).
We use this reference because AI-supported decisions must be explained under specific circumstances according to the GDPR and the AI Act~\cite{panigutti_role_2023}.

In their words: ``A model is considered to be transparent if by itself it is understandable.''~\cite{arrieta_explainable_2019}.
Such models include linear regression, decision trees, k-nearest neighbors, rule-based learning, general additive models, and Bayesian models.
Nonetheless, models that are not deemed intrinsically transparent can be made explainable through the use of \textit{post-hoc} techniques.
Neural networks (especially deep and convoluted) and Support Vector Machines (SVM) typically fall under this category, as well as reinforcement learning~\cite{puiutta_explainable_2020}.

\section{Methodology}
\label{sec:methodology} 
This study adopts the widely known methodology for systematic literature reviews (SLRs) proposed by Kitchenham~\cite{kitchenham2004procedures}. The SLR methodology offers us a well-defined and rigorous sequence of methodological steps consisting of three main phases: (1) planning, (2) conducting, and (3) reporting the review.
An SLR Protocol that describes the entire research process has been written for this study.
% (a summary version of which can be found in Appendix~\ref{app:protocol}). 
Furthermore, we make our research data openly available in a GitHub repository for reproducibility.\footnote{\url{https://github.com/Victor-Morel/SLR_AI_PPA}} 
Our material comprises the citation files of each query, the Data Extraction Forms (DEFs) of the selected papers, and the charting spreadsheet used to compile our data. 
We refer to these documents for methodological details.

\subsection{Planning the Review}\label{sec:slr-planning}
The first activity of the planning phase was to determine the need for this SLR. Several databases were searched to verify if any surveys or reviews had been conducted on \PPAs. Search terms such as privacy, data protection, assistant, agent, artificial intelligence, and machine learning were used. However, we could not identify any survey or systematic reviews on the topic, reassuring the need for an SLR.
The research questions, presented in Section~\ref{sec:introduction}, guided the remaining phases of this SLR with respect to the search process, selection criteria, and data synthesis.

\subsection{Conducting the Review}\label{sec:slr-conducting}
\subsubsection{Search Strategy}
Based on our RQs and previous preliminary searches when designing the SLR Protocol, we identified a list of nine relevant keywords, i.e., \textit{privacy, data protection, assistant, agent, artificial intelligence, machine learning, intelligent, automatic}, and \textit{personalized}. These keywords were used to construct the search query in Listing \ref{lst:search-query}. As such, the search query targets papers working on three joint topics: 1) privacy (\textit{or} data protection), using either 2) an assistant \textit{or} an agent, \textit{and} leveraging 3) artificial intelligence \textit{or} personalization. 

\begin{center}
  \lstset{%
    caption=Composition of Search Query for Literature Search.,
    basicstyle=\ttfamily\footnotesize\bfseries,
    frame=tb,
    backgroundcolor=\color{lightgray},
    label={lst:search-query},
    captionpos=b
  }
  \begin{lstlisting}
    Search Query = {
      (privacy OR "data protection") AND 
      (assistant* OR agent*) AND ("artificial 
      intelligence" OR "machine*learning" 
      OR intelligent OR automat* OR personali*ed)
    }
  \end{lstlisting}
\end{center}

%\begin{spverbatim}
%(privacy OR "data protection") AND (assistant* OR agent*) AND ("artificial intelligence" OR "machine*learning" OR intelligent OR automat* OR personali*ed)
%\end{spverbatim} 

Four scientific databases were selected, i.e., Scopus, Web of Science, IEEE Xplore, and ACM Digital Library, due to their high relevance to the areas of computer science and engineering, comprising the vast majority of published research in the field.
We also specified inclusion and exclusion criteria (see Table \ref{tab:selection-criteria}) used during the screening of publications retrieved from the databases.
Marky et al.~\cite{marky_decide_2024} is a good illustration of a relevant paper not meeting our selection criteria.
In spite of providing a technical solution to automate privacy decisions (explicitly called a PPA), the paper does not use AI and hence is not included in our list of papers.
Bollinger et al.~\cite{bollinger_automating_2022} provides another example of an excluded yet relevant paper.
The paper provides a technical solution for automating privacy decisions and uses AI, but does not personalize the decisions.
% \todo[inline]{discuss the CHI2024 paper as an example of excluded (yet relevant) paper}

\begin{table}[!h]
    \centering
    \caption{Criteria for the inclusion and exclusion of studies.}
    \label{tab:selection-criteria}
    \small
    \begin{tabular}{|p{0.95\linewidth}|}
    \hline
       \textbf{Inclusion Criteria} \\ \hline
       - Provides a technical solution (implemented or theoretical) to help end-users automate personal (and personalized) privacy decisions with an assistant (or artificial agent) in IT systems. \\
       - Papers from 2013 onward to concentrate on the state-of-the-art. \\
       - The concept of AI needs to be explicitly stated in the papers. \\ \hline \hline
       \textbf{Exclusion Criteria} \\ \hline
       - Papers with solutions that are purely theoretical without substantial explanations on how they could be implemented in practice. \\
       - Papers with solutions that solely automate the analysis of privacy policies but without any type of personalization. \\
       - Papers with poor scientific quality (e.g., lack objectives or research questions, the methodology is not described, the solution is insufficiently/vaguely described, etc.). \\
       \hline
    \end{tabular}
\end{table}

Before starting the search process, two authors piloted the searches on all databases and ran a \textit{calibration exercise} to verify the consistency of the inclusion criteria. For that, the authors independently screened 10\% of the results and discussed their decisions. The conflicts were all discussed and solved, sometimes with the help of the third author. This process was repeated a second time, screening another 10\% of the papers at a point that the authors fully agreed with the consistency of the selection process.

% \subsubsection{Study Records}
\subsubsection{Data management}
To manage the screening process, we exported search results from each database and imported them to the RAYYAN software (\url{https://rayyan.ai/}), allowing two reviewers to independently select papers (i.e., double-blinded) and to manage conflicts by a third reviewer. 
Duplicated publications were also removed using RAYYAN during the selection process. 
Bibliographies of final results were exported to Zotero (for citing and sharing research).

\subsubsection{Selection Process}
% \todo[inline]{Selection issues should be fixed by the update}

\begin{figure*}
    \centering
    \includegraphics[scale=.2]{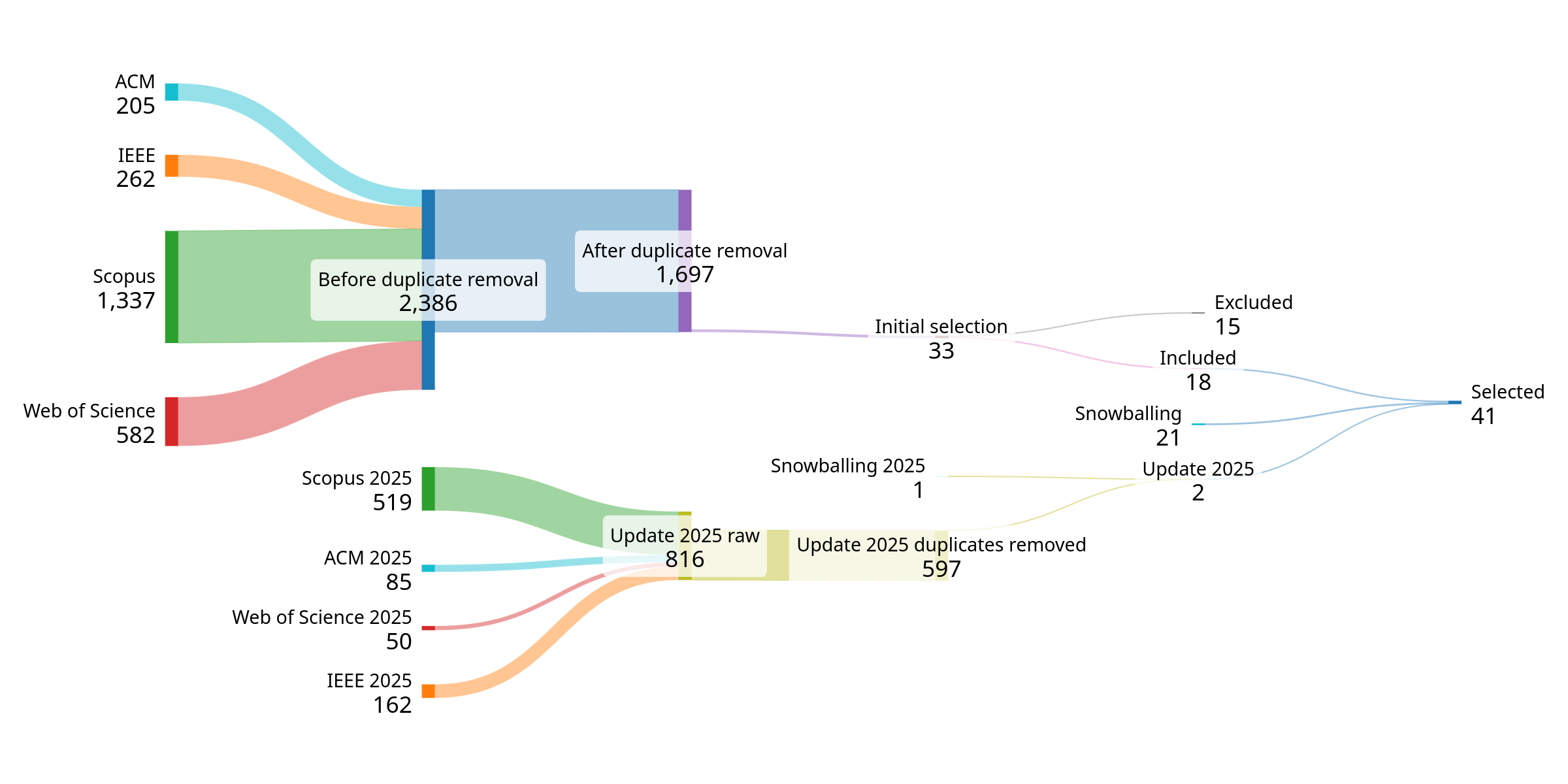}
    \caption{Sankey chart of the selection process.}
    \label{fig:sankey}
\end{figure*}

Figure~\ref{fig:sankey} presents an overview of the selection process.
The querying of the databases mentioned above on October 19, 2023, yielded 2386 papers and 1697 unique entries after removing duplicates.
The screening phase lasted until November 23, 2023, and resulted in the selection of 33 papers.
Two authors then read 10\% of these 33 papers (3) and adjusted the DEF based on mutual feedback. This step helped us add new important fields to the DEF and consistently extract data from the papers.

The first data extraction phase, consisting of a full reading of each of the 33 papers, was performed over weeks 4 to 7 (included) in 2024.
Fifteen papers were excluded after full reading for different reasons: they were duplicates (i.e., same work published in different venues); they did not provide any technical solution; the automated decisions were not personalized to an end user; AI was not used for automating decisions; or they are of poor scientific quality (see our criteria in Table~\ref{tab:selection-criteria}); one paper was not available for download, we could not access it even after reaching out the authors.

We then proceeded to several snowballing phases~\cite{wohlin_guidelines_2014}, during which we checked the abstracts of all seemingly relevant\footnote{We only assessed papers cited in relevant sections, e.g., related work.} papers cited (backward snowballing), and screened citing papers (forward snowballing).
The snowballing process lasted from week 8 to week 19 of 2024 and resulted in 21 additional papers after exclusion, for a total of 39 papers (33-15+21).
We performed an update of our results in February 2025, repeating the whole process in a 2nd round of searches. This update led us to include one additional paper through the database searches and another one through the snowballing process, for a total of 41 papers.

\subsubsection{Data Extraction and Analysis}
\label{subsec:analysis}
% \todo[inline]{More details on the methodology → @victor (after redoing the critical appraisal)}
% \todo[inline]{double-check consistency after merging appendix}

% \paragraph{Data extraction}
The data extracted in the DEFs was compiled and further organized in spreadsheets during weeks 20-21 in 2024.
This process also included the initial aggregation of data and the creation of frequency charts across several data categories (e.g., studies per year, types of publications, authors and affiliations, etc.). Table \ref{tab:def-summary} shows a summary of the components extracted from the included studies.
%\begin{itemize}
%    \item Bibliographic information, such as title, abstract, authors and affiliations, venues, year of publication, etc.
%    \item Key information of the \PPA, such as its source(s) of data, its eventual architecture, its system context, the type of privacy decision considered, the accuracy of the decisions, the type of AI used, etc.
%    \item The presence of a user study, and performed its critical appraisal (see below)
%    \item Extent of evaluation – scale of validation activity that is measured
%    \item Quality assessment and critical appraisal of the studies that have validated or evaluated the \PPA
%    \item Features of users control over decisions (initially guided by EU consent requirements)
%\end{itemize}

\begin{table}[!h]
    \centering
    \caption{Summary of information included in the DEFs.}
    \label{tab:def-summary}
    \small
    \begin{tabular}{|p{0.95\linewidth}|}
    \hline
       \textbf{Extracted Information} \\ \hline
       - Bibliographic information, such as title, abstract, authors and affiliations, venues, year of publication, etc. \\
       - Key information of the \PPA, such as its source(s) of data, its eventual architecture, its system context, the type of privacy decision considered, the accuracy of the decisions, the type of AI used, etc. \\
       - The presence of a user study, and performed its critical appraisal (see below). \\
       - Extent of evaluation, a scale of validation activity that is measured. \\
       - Quality assessment and critical appraisal of the studies that have validated or evaluated the \PPA. \\
       - Features of user's control over decisions (initially guided by EU consent requirements). \\
    \hline
    \end{tabular}
\end{table}

% \paragraph{Types of contributions}
It is also worth noting that we classified publications by their types of contributions according to the following items proposed by Kuhrmann et al.~\cite{kuhrmann_software_2016} and Shaw~\cite{shaw_writing_2003} (i.e., all that apply): \emph{(i) model}, as a representation of observed reality by concepts after conceptualization; \emph{(ii) theory}, as a construct of cause-effect relationships; \emph{(iii) framework}, including frameworks/methods (related to automated privacy decisions); \emph{(iv) guidelines}, as a list of advice; \emph{(v) lessons learned}, as a set of outcomes from obtained results; \emph{(vi) advice}, as recommendations (from opinion); and, \emph{(vii) tool}, as tools to automate privacy decisions.

% \usepackage{multirow}
\begin{table}[]
\caption{This table informs the \textbf{type of contribution}, informing whether the surveyed solution presents a \textit{framework}, a \textit{tool} (i.e. with an implementation), a \textit{model}, \textit{lessons learned}, or \textit{advice}.}
\label{tab:contribution}
\begin{tabular}{l|l|llllll}
\hline
\cellcolor[HTML]{FFFFFF} &  & \multicolumn{6}{l}{Type of contribution} \\ \cline{3-8} 
\cellcolor[HTML]{FFFFFF} &  &  &  &  &  &  &  \\
\cellcolor[HTML]{FFFFFF} &  &  &  &  &  &  &  \\
\cellcolor[HTML]{FFFFFF} &  &  &  &  &  &  &  \\
\cellcolor[HTML]{FFFFFF} &  &  &  &  &  &  &  \\
\cellcolor[HTML]{FFFFFF} &  &  &  &  &  &  &  \\
\cellcolor[HTML]{FFFFFF} &  &  &  &  &  &  &  \\
\multirow{-8}{*}{\cellcolor[HTML]{FFFFFF}Year} & \multirow{-8}{*}{Publication} & \begin{rotate}{90} Framework \end{rotate} & \begin{rotate}{90} Tool \end{rotate} & \begin{rotate}{90} Model \end{rotate} & \begin{rotate}{90} Theory \end{rotate} & \begin{rotate}{90} Lessons learned \end{rotate} & \begin{rotate}{90} Advice \end{rotate} \\ \hline
2014 & Xie et al.~\cite{xie_location_2014} & $\bullet$ &  &  &  & $\bullet$ &  \\
\rowcolor[HTML]{C0C0C0} 
2015 & Apolinarski et al.~\cite{apolinarski_automating_2015} & $\bullet$ & $\bullet$ &  &  &  &  \\
2015 & Hirschprung et al.~\cite{hirschprung_simplifying_2015} & $\bullet$ &  & $\bullet$ &  &  &  \\
\rowcolor[HTML]{C0C0C0} 
2015 & Squicciarini et al.~\cite{squicciarini_privacy_2015} & $\bullet$ &  & $\bullet$ &  &  &  \\
2016 & Liu et al.~\cite{liu_follow_2016} &  & $\bullet$ &  &  & $\bullet$ &  \\
\rowcolor[HTML]{C0C0C0} 
2016 & Albertini et al.~\cite{albertini_privacy_2016} &  & $\bullet$ &  &  &  &  \\
2016 & Dong et al.~\cite{dong_ppm_2016} &  &  & $\bullet$ &  & $\bullet$ &  \\
\rowcolor[HTML]{C0C0C0} 
2017 & Baarslag et al.~\cite{baarslag_automated_2017} &  & $\bullet$ & $\bullet$ &  & $\bullet$ &  \\
2017 & Fogues et al.~\cite{fogues_sosharp_2017} &  & $\bullet$ &  &  &  &  \\
\rowcolor[HTML]{C0C0C0} 
2017 & Zhong et al.~\cite{zhong_group-based_2017} &  &  & $\bullet$ &  &  &  \\
2017 & Misra et al.~\cite{misra_pacman_2017} &  & $\bullet$ &  &  &  &  \\
\rowcolor[HTML]{C0C0C0} 
2017 & Nakamura et al.~\cite{camp_easing_2017} &  &  & $\bullet$ &  &  &  \\
2017 & Olejnik et al.~\cite{olejnik_smarper_2017} & $\bullet$ & $\bullet$ &  &  &  &  \\
\rowcolor[HTML]{C0C0C0} 
2018 & Das et al.\cite{das_personalized_2018} &  & $\bullet$ &  &  &  &  \\
2018 & Tan et al.~\cite{tan_context-perceptual_2018} &  & $\bullet$ &  &  &  &  \\
\rowcolor[HTML]{C0C0C0} 
2018 & Wijesekera et al.~\cite{wijesekera_contextualizing_2018} &  & $\bullet$ &  &  & $\bullet$ &  \\
2018 & Yu et al.~\cite{yu_leveraging_2018} &  &  & $\bullet$ & $\bullet$ &  &  \\
\rowcolor[HTML]{C0C0C0} 
2018 & Bahirat et al.~\cite{bahirat_data-driven_2018} &  &  & $\bullet$ &  &  &  \\
2018 & Raber et al.~\cite{raber_retailio_2018} &  & $\bullet$ & $\bullet$ &  & $\bullet$ &  \\
\rowcolor[HTML]{C0C0C0} 
2019 & Klingensmith et al.~\cite{klingensmith_hypervisor-based_2019} &  & $\bullet$ &  &  &  &  \\
2019 & Barbosa et al.~\cite{barbosa_what_2019} &  &  & $\bullet$ &  &  & $\bullet$ \\
\rowcolor[HTML]{C0C0C0} 
2019 & Alom et al.~\cite{alom_helping_2019} & $\bullet$ &  &  &  &  &  \\
2019 & Alom et al.~\cite{alom_adapting_2019} & $\bullet$ &  &  &  &  &  \\
\rowcolor[HTML]{C0C0C0} 
2020 & Kasaraneni et al.~\cite{barolli_selflearning_2020} &  & $\bullet$ & $\bullet$ &  &  &  \\
2020 & Kaur et al.~\cite{kaur_smart_2020} &  &  & $\bullet$ &  &  &  \\
\rowcolor[HTML]{C0C0C0} 
2020 & Botti-Cebria et al.~\cite{herrero_automatic_2021} &  & $\bullet$ &  &  &  &  \\
2020 & Kökciyan et al.~\cite{kokciyan_turp_2020} &  &  & $\bullet$ &  &  &  \\
\rowcolor[HTML]{C0C0C0} 
2020 & Sanchez et al.~\cite{sanchez_recommendation_2020} &  &  & $\bullet$ &  &  &  \\
2021 & Kaur et al.~\cite{barolli_reinforcement_2021} &  &  & $\bullet$ &  &  &  \\
\rowcolor[HTML]{C0C0C0} 
2021 & Lobner et al.~\cite{lobner_explainable_2021} &  &  & $\bullet$ &  &  &  \\
2022 & Filipczuk et al.~\cite{filipczuk_automated_2022} & $\bullet$ & $\bullet$ &  &  &  &  \\
\rowcolor[HTML]{C0C0C0} 
2022 & Hirschprung et al.~\cite{hirschprung_game_2022} & $\bullet$ &  & $\bullet$ &  &  &  \\
2022 & Kökciyan et al.~\cite{kokciyan_taking_2022} &  &  & $\bullet$ &  &  &  \\
\rowcolor[HTML]{C0C0C0} 
2022 & Ulusoy et al.~\cite{ulusoy_panola_2022} &  &  & $\bullet$ &  &  &  \\
2022 & Zhan et al.~\cite{zhan_model_2022} &  &  & $\bullet$ &  &  &  \\
\rowcolor[HTML]{C0C0C0} 
2022 & Brandão et al.~\cite{brandao_prediction_2022} &  & $\bullet$ &  &  &  &  \\
2022 & Mendes et al.~\cite{mendes_enhancing_2022} &  & $\bullet$ & $\bullet$ &  & $\bullet$ &  \\
\rowcolor[HTML]{C0C0C0} 
2022 & Shanmugarasa et al.~\cite{shanmugarasa_automated_2022} &  & $\bullet$ & $\bullet$ &  &  &  \\
2023 & Ayci et al.~\cite{ayci_uncertainty-aware_2023} &  & $\bullet$ & $\bullet$ &  &  &  \\
\rowcolor[HTML]{C0C0C0} 
2023 & Serramia et al.~\cite{serramia_predicting_2023} &  & $\bullet$ & $\bullet$ &  &  &  \\
2024 & Wang et al.~\cite{wang_privacyoracle_2024} &  & $\bullet$ &  &  &  &  \\ \hline
\end{tabular}
\end{table}

% \paragraph{Data synthesis}
% We collated and summarized results into classification tables.
% We composed a narrative synthesis for papers meeting our inclusion criteria.

% Preliminary components of the data synthesis:
% \begin{itemize}
%     \item Overall identification and classification of \PPAs in published research
%     \item Classification tables presenting features of \PPAs used to coherently organize the solutions surveyed
%     \item Comparison analysis -- based on the features of \PPAs --, and narrative synthesis
% \end{itemize}
Although we attempted to extract as much data from the studies as possible using a DEF, we found that, during the data analysis process, there was a need to further categorize studies across other \textit{facets}.
For instance, additional information was compiled in the spreadsheets, such as a high-level categorization of certain fields (i.e., the type of AI used) or a critical appraisal of the user studies presented in the selected papers, although only when those user studies were used to evaluate the \PPA, and not when they were used for data collection to build datasets.

This collection of facets created during the study design and data analysis processes forms the basis of the work's final classification scheme, presented as part of the main results.
All authors were involved in the data analysis process and the definition of facets that further classify studies on the topic.

%\paragraph{Critical Appraisal}
\label{subsec:appraisal}
% Lastly, we also performed a critical appraisal on the user studies presented, although only when those user studies were used to evaluate the \PPA, and not when they were used for data collection for dataset building.
For the \textit{critical appraisal}, we used the CAT (Critically Appraised Topic) Manager App of CEBMA (the Center for Evidence-Based Management)~\cite{cebma_cat_2025}, which provides a practical yet rigorous approach to evaluate studies based on objective criteria.
It helps determine a study's trustworthiness regarding cause and effect questions.
Once a study is evaluated, the possible outcomes are: Very high (A+), High (A), Moderate (B), Limited (C), Low (D), or Very low (D-).

By definition, an \PPA leverages AI techniques. We therefore collected information about the \textit{Type of AI used}.
AI models rely on data for training and decision-making. As such, we extracted the \textit{source of data}.
During the adjustment of the DEF, we observed that \PPAs are usually designed for a specific \textit{system context}, and for one or several \textit{types of decision}.
Connected to the system context, we extracted the \textit{choice of architecture} of the implementation (if any) to analyze the trust implications.

We also collected the methods for an \textit{empirical assessment}, presence, and quality of user studies, or the means used to measure the accuracy, to gain insights on eventual benchmarks of \PPAs.
Studies can be classified as evaluation or validation research, as proposed by Wieringa et al.~\cite{wieringa_requirements_2006}. 
An evaluation works in real-world practice and is implementing/deploying the solution or testing in an actual project with real test users, such as real case studies and realistic user testing of prototypes/systems.
A validation is a limited illustrative or hypothetical ``case study'' or ``use case'' performed as a lab experiment.
In general practice, prototypes are often validated by cross-sectional studies.

Finally, initially guided by legal requirements for consent and the exercise of data subject rights under the GDPR (although eventually, no paper considers consent), we extracted what became \textit{user control over decisions}.

\subsection{Reporting the Review}
Based on the data analysis, a whole coherent narrative was written by the research team, i.e., this SLR article, conveying all the results, our interpretation of the main findings, and identifying research gaps.
This synthesis on \PPAs is thus reported in the following sections.

\section{Summary of Data Charting Results}
\label{sec:results}
% \todo[inline]{update table references}
This section provides an overview of the quantitative insights generated through the data charting process (e.g., publications per year, citations, types of decisions).
The main findings related to the critical appraisals are also introduced in this section.
It is worth noting, nonetheless, that the classification features are further detailed in the following Section \ref{sec:classification}.
% , except for the \textit{Empirical assessment} which, along with the results of our critical appraisal and the type of contribution, are reported in Table~\ref{tab:appendix_table} due to space constraints.

Among the 41 papers surveyed, we tallied 15 different countries for the authors' affiliations (see Table~\ref{tab:countries}), with the USA and UK leading in numbers.
About 55\% ($n=22$) of the selected publications were published from 2019 to 2024, with the year 2021 being the most productive with 8 publications (see Table~\ref{tab:years}).
At the time of data collection, papers were cited between 0 and 275 times with an average of 38.56 citations, a median of 14, and a standard deviation of 59, indicating a power law distribution of the citation count.
The most cited papers are Liu et al.~\cite{liu_follow_2016} ($n=275$), Yu et al.~\cite{yu_leveraging_2018} ($n=199$), and Squicciarini et al.~\cite{squicciarini_privacy_2015} ($n=130$) (numbers at the time of data collection).

\begin{table}[h]
\centering
\caption{Countries of affiliation of authors of selected papers.}
\label{tab:countries}
\begin{tabular}{@{}lc||lc@{}}
\toprule
\textbf{Countries} & \textbf{Total} & \textbf{Countries} & \textbf{Total} \\ \midrule
United States & 14 & China & 2 \\
United Kingdom & 6 & Israel & 2 \\
Japan & 4 & Portugal & 2 \\
Netherlands & 4 & Switzerland & 1 \\
Italy & 4 & Turkey & 1 \\
Germany & 4 & Canada & 1 \\
Spain & 3 & Australia & 1 \\
India & 2 & & \\ \bottomrule
\end{tabular}
\end{table}

\begin{table}[h]
\centering
\caption{Number of publications per year.}
\label{tab:years}
\begin{tabular}{@{}lc||lc@{}}
\toprule
\textbf{Year} & \textbf{N. of Publications} & \textbf{Year} & \textbf{N. of Publications} \\ \midrule
2013 & 0 & 2019 & 4 \\
2014 & 1 & 2020 & 5 \\
2015 & 3 & 2021 & 2 \\
2016 & 3 & 2022 & 8 \\
2017 & 6 & 2023 & 2 \\
2018 & 5 & 2024 & 1 \\ \bottomrule
\end{tabular}
\end{table}

As shown in Table \ref{tab:classification_1}, regarding the sources of data used by the \PPAs, context data, attitudinal data, and metadata were the most prevalent.
We observed a relatively balanced distribution when it comes to the types of decisions (between 12 and 15 for each type) and the system contexts (between 11 and 13, with two outliers for \textit{Cloud} and \textit{Intelligent retail store}).
For the types of AI systems that we were able to classify, most models were deemed non-intrinsically transparent (NIT, $n=14$), followed by transparent (T, $n=8$) and partially transparent (PT, $n=4$) models.
Note also that Das et al.~\cite{das_personalized_2018} did not specify the type of AI used in their paper, we were therefore unable to categorize their solution in that respect (under \textit{Type of AI used}).

In Table \ref{tab:contribution}, the publications were also classified by their types of contributions, according to the categories proposed by Kuhrmann et al.~\cite{kuhrmann_software_2016} and Shaw~\cite{shaw_writing_2003}.
% (see Appendix \ref{app:protocol}).
We observed a prevalence of models ($n=24$) and tools ($n=21$), followed by frameworks ($n=9$).
Nonetheless, these models, tools, and frameworks lack empirical assessment, an issue further analyzed in Section \ref{subsec:taxo_validation}.

Finally, the results of our critical appraisal can be found in Table~\ref{tab:classification_2}.
% in the Appendix \ref{app:tables}.
Out of the 41 publications, only 15 presented a user study, i.e., qualitative research that is suitable to be critically appraised. 
In terms of quality, they mostly scored ``low'' or ``very low'' ($n=9$) according to the CEBMA checklist.
Exceptionally, only the studies of Liu et al.~\cite{liu_follow_2016} and Baarslag et al.~\cite{baarslag_automated_2017} were appraised as of high quality.
This suggests that the empirical evidence around \PPAs remains incipient, and existing solutions can be further tested in real-world settings, a challenge (or opportunity) that is discussed in Section \ref{sec:discussion}.

\section{Classification for \PPAs}
\label{sec:classification}
\begin{figure*}
    \centering
    \includegraphics[scale=.5]{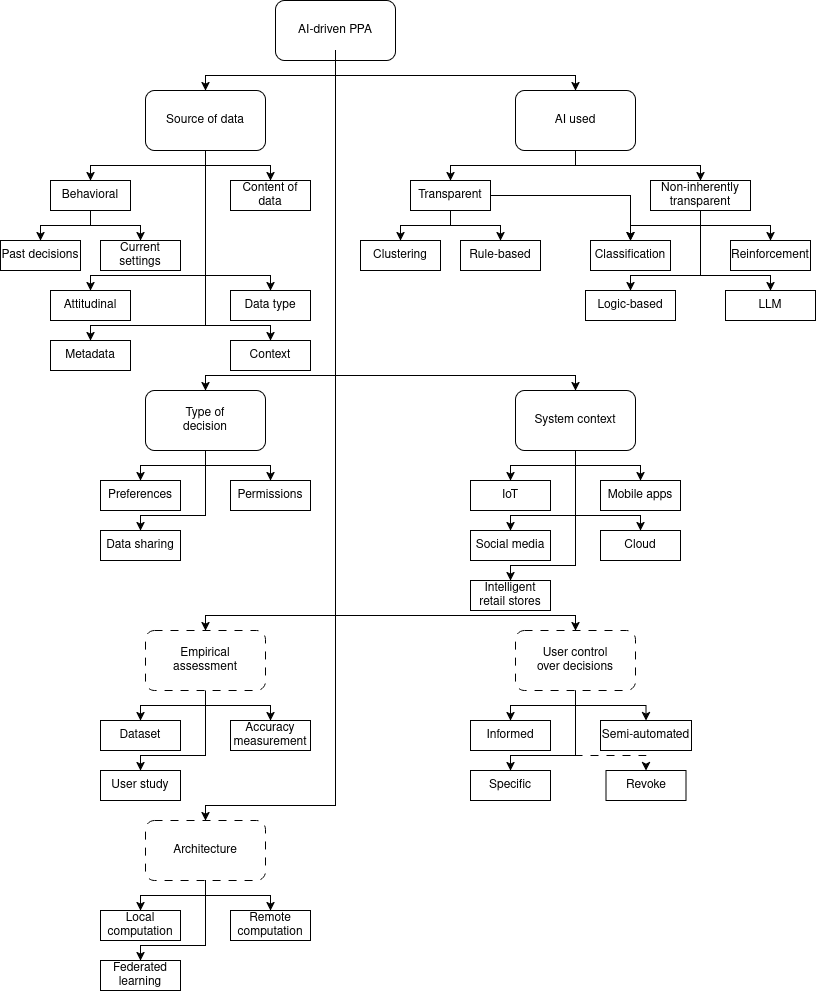}
    \caption{A schematic representation of the classification presented in Section~\ref{sec:classification}. 
    Each facet is represented as a \textit{rounded} box, \textit{solid} for the mandatory features, and \textit{dashed} for the optional ones.
    For \textbf{User control over decisions} (see Section~\ref{subsec:taxo_req}), we distinguish between \textit{qualities} of control (solid arrows) and \textit{instruments} of control (dashed arrows).}
    \label{fig:classification}
\end{figure*}

We provide in this section a classification for \PPAs as the main contribution of this SLR.
Summarized in Figure \ref{fig:classification}, the classification comprises several dimensions, i.e., features typically considered in the design of such an assistant (see also Tables \ref{tab:classification_1} and \ref{tab:classification_2}).
These dimensions are the \textit{type of decision} (Section~\ref{subsec:taxo_decision}), the \textit{type of AI} (Section~\ref{subsec:taxo_ai}) and the \textit{source of data} (Section~\ref{subsec:taxo_source}) used in the decision, the \textit{system context} (Section~\ref{subsec:taxo_system_context}), the \textit{choice architecture} of its eventual implementation (Section~\ref{subsec:taxo_architecture}), the \textit{empirical assessment} (Section~\ref{subsec:taxo_validation}), and the extent to which \textit{users have control over the decisions} (Section~\ref{subsec:taxo_req}).

\begin{table*}[]
\centering
\caption{Summary table of our classification, part 1.
~
It presents the mandatory features of \PPAs, namely the \textbf{type of decision}, the \textbf{type of AI used} (for which we specified whether the classification model is Transparent (T), Not-Inherently Transparent (NIT), or Partially Transparent (PT) because several models are used), the \textbf{type of source of data}, and the \textbf{system context} (note that IRS stands for Intelligent Retail Store).
~
An empty field signifies that the solution does not exhibit the characteristic (e.g., does not consider Y type of decision).}
\label{tab:classification_1}
\begin{tabular}{l|l|lll|llllll|llllll|lllll}
\hline
 &  & \multicolumn{3}{l|}{Type of decision} & \multicolumn{6}{l|}{Type of AI used} & \multicolumn{6}{l|}{Type of source of data} & \multicolumn{5}{l}{System context} \\ \cline{3-22} 
 &  &  &  &  &  &  &  &  &  &  &  &  &  &  &  &  &  &  &  &  &  \\
 &  &  &  &  &  &  &  &  &  &  &  &  &  &  &  &  &  &  &  &  &  \\
 &  &  &  &  &  &  &  &  &  &  &  &  &  &  &  &  &  &  &  &  &  \\
 &  &  &  &  &  &  &  &  &  &  &  &  &  &  &  &  &  &  &  &  &  \\
 &  &  &  &  &  &  &  &  &  &  &  &  &  &  &  &  &  &  &  &  &  \\
 &  &  &  &  &  &  &  &  &  &  &  &  &  &  &  &  &  &  &  &  &  \\
\multirow{-8}{*}{Year} & \multirow{-8}{*}{Publication} & \cellcolor[HTML]{FFFFFF}\begin{rotate}{90} Permissions \end{rotate} & \cellcolor[HTML]{FFFFFF}\begin{rotate}{90} Preferences \end{rotate} & \cellcolor[HTML]{FFFFFF}\begin{rotate}{90} Data sharing \end{rotate} & \cellcolor[HTML]{FFFFFF}\begin{rotate}{90} Classification \end{rotate} & \cellcolor[HTML]{FFFFFF}\begin{rotate}{90} Clustering \end{rotate} & \cellcolor[HTML]{FFFFFF}\begin{rotate}{90} Rule-based \end{rotate} & \cellcolor[HTML]{FFFFFF}\begin{rotate}{90} Logic-based \end{rotate} & \cellcolor[HTML]{FFFFFF}\begin{rotate}{90} Reinforcement \end{rotate} & \cellcolor[HTML]{FFFFFF}\begin{rotate}{90} LLM \end{rotate} & \cellcolor[HTML]{FFFFFF}\begin{rotate}{90} Context \end{rotate} & \cellcolor[HTML]{FFFFFF}\begin{rotate}{90} Attitudinal data \end{rotate} & \cellcolor[HTML]{FFFFFF}\begin{rotate}{90} Metadata \end{rotate} & \cellcolor[HTML]{FFFFFF}\begin{rotate}{90} Data type \end{rotate} & \cellcolor[HTML]{FFFFFF}\begin{rotate}{90} Content of data \end{rotate} & \cellcolor[HTML]{FFFFFF}\begin{rotate}{90} Behavioral data \end{rotate} & \cellcolor[HTML]{FFFFFF}\begin{rotate}{90} Mobile apps \end{rotate} & \cellcolor[HTML]{FFFFFF}\begin{rotate}{90} Social media \end{rotate} & \cellcolor[HTML]{FFFFFF}\begin{rotate}{90} IoT \end{rotate} & \cellcolor[HTML]{FFFFFF}\begin{rotate}{90} Cloud \end{rotate} & \cellcolor[HTML]{FFFFFF}\begin{rotate}{90} IRS \end{rotate} \\ \hline
2014 & Xie et al.~\cite{xie_location_2014} &  & $\bullet$ &  & NIT &  &  &  &  &  & $\bullet$ &  &  &  &  &  &  &  &  &  &  \\
\rowcolor[HTML]{C0C0C0} 
2015 & Apolinarski et al.~\cite{apolinarski_automating_2015} & $\bullet$ &  &  & NIT &  &  &  &  &  & $\bullet$ &  &  &  &  &  & $\bullet$ &  &  &  &  \\
2015 & Hirschprung et al.~\cite{hirschprung_simplifying_2015} & $\bullet$ &  &  &  & $\bullet$ &  &  &  &  &  & $\bullet$ &  &  &  &  &  &  &  & $\bullet$ &  \\
\rowcolor[HTML]{C0C0C0} 
2015 & Squicciarini et al.~\cite{squicciarini_privacy_2015} &  &  & $\bullet$ &  &  & $\bullet$ &  &  &  & $\bullet$ &  & $\bullet$ &  & $\bullet$ &  &  &  &  &  &  \\
2016 & Liu et al.~\cite{liu_follow_2016} & $\bullet$ &  &  & T & $\bullet$ &  &  &  &  &  & $\bullet$ &  &  &  &  & $\bullet$ &  &  &  &  \\
\rowcolor[HTML]{C0C0C0} 
2016 & Albertini et al.~\cite{albertini_privacy_2016} &  &  & $\bullet$ &  &  & $\bullet$ &  &  &  &  & $\bullet$ &  &  &  &  &  & $\bullet$ &  &  &  \\
2016 & Dong et al.~\cite{dong_ppm_2016} &  &  & $\bullet$ & T &  &  &  &  &  & $\bullet$ &  &  &  & $\bullet$ &  &  & $\bullet$ &  &  &  \\
\rowcolor[HTML]{C0C0C0} 
2017 & Baarslag et al.~\cite{baarslag_automated_2017} & $\bullet$ &  &  &  &  &  & $\bullet$ &  &  &  & $\bullet$ &  & $\bullet$ &  &  & $\bullet$ &  &  &  &  \\
2017 & Fogues et al.~\cite{fogues_sosharp_2017} &  &  & $\bullet$ & PT &  &  &  &  &  & $\bullet$ & $\bullet$ &  &  &  &  &  & $\bullet$ &  &  &  \\
\rowcolor[HTML]{C0C0C0} 
2017 & Zhong et al.~\cite{zhong_group-based_2017} &  &  & $\bullet$ & NIT &  &  &  &  &  &  & $\bullet$ &  & $\bullet$ &  &  &  & $\bullet$ &  &  &  \\
2017 & Misra et al.~\cite{misra_pacman_2017} &  &  & $\bullet$ & NIT &  &  &  &  &  & $\bullet$ &  &  &  & $\bullet$ &  &  & $\bullet$ &  &  &  \\
\rowcolor[HTML]{C0C0C0} 
2017 & Nakamura et al.~\cite{camp_easing_2017} &  & $\bullet$ &  & NIT &  &  &  &  &  &  &  &  & $\bullet$ &  & $\bullet$ &  &  &  &  &  \\
2017 & Olejnik et al.~\cite{olejnik_smarper_2017} & $\bullet$ &  &  & T &  &  &  &  &  & $\bullet$ & $\bullet$ &  &  &  &  & $\bullet$ &  &  &  &  \\
\rowcolor[HTML]{C0C0C0} 
2018 & Das et al.\cite{das_personalized_2018} &  & $\bullet$ &  &  &  &  &  &  &  &  & $\bullet$ &  &  &  &  &  &  & $\bullet$ &  &  \\
2018 & Tan et al.~\cite{tan_context-perceptual_2018} & $\bullet$ &  &  & T &  &  &  &  &  &  &  & $\bullet$ &  &  &  & $\bullet$ &  &  &  &  \\
\rowcolor[HTML]{C0C0C0} 
2018 & Wijesekera et al.~\cite{wijesekera_contextualizing_2018} & $\bullet$ &  &  & NIT &  &  &  &  &  & $\bullet$ &  & $\bullet$ & $\bullet$ &  &  & $\bullet$ &  &  &  &  \\
2018 & Yu et al.~\cite{yu_leveraging_2018} &  &  & $\bullet$ & NIT &  &  &  &  &  & $\bullet$ &  &  &  & $\bullet$ &  &  & $\bullet$ &  &  &  \\
\rowcolor[HTML]{C0C0C0} 
2018 & Bahirat et al.~\cite{bahirat_data-driven_2018} &  & $\bullet$ &  & T &  &  &  &  &  &  & $\bullet$ &  &  &  &  &  &  & $\bullet$ &  &  \\
2018 & Raber et al.~\cite{raber_retailio_2018} & $\bullet$ &  &  & T &  &  &  &  &  &  & $\bullet$ &  &  &  & $\bullet$ &  &  &  &  & $\bullet$ \\
\rowcolor[HTML]{C0C0C0} 
2019 & Klingensmith et al.~\cite{klingensmith_hypervisor-based_2019} & $\bullet$ &  &  & NIT &  &  &  &  &  &  &  & $\bullet$ &  &  & $\bullet$ &  &  & $\bullet$ &  &  \\
2019 & Barbosa et al.~\cite{barbosa_what_2019} &  & $\bullet$ &  & PT &  &  &  &  &  &  & $\bullet$ & $\bullet$ &  &  &  &  &  & $\bullet$ &  &  \\
\rowcolor[HTML]{C0C0C0} 
2019 & Alom et al.~\cite{alom_helping_2019} &  & $\bullet$ &  & PT &  &  &  &  &  & $\bullet$ & $\bullet$ &  &  &  &  &  &  &  &  &  \\
2019 & Alom et al.~\cite{alom_adapting_2019} &  & $\bullet$ &  & NIT &  &  &  &  &  &  & $\bullet$ &  &  &  & $\bullet$ &  &  & $\bullet$ &  &  \\
\rowcolor[HTML]{C0C0C0} 
2020 & Kasaraneni et al.~\cite{barolli_selflearning_2020} &  &  & $\bullet$ & T & $\bullet$ &  &  &  &  &  &  & $\bullet$ &  &  &  &  & $\bullet$ &  &  &  \\
2020 & Kaur et al.~\cite{kaur_smart_2020} &  & $\bullet$ &  & NIT &  &  &  &  &  & $\bullet$ &  & $\bullet$ &  &  &  & $\bullet$ &  & $\bullet$ &  &  \\
\rowcolor[HTML]{C0C0C0} 
2020 & Botti-Cebria et al.~\cite{herrero_automatic_2021} &  &  & $\bullet$ & PT &  &  &  &  &  &  &  &  &  & $\bullet$ &  &  & $\bullet$ &  &  &  \\
2020 & Kökciyan et al.~\cite{kokciyan_turp_2020} & $\bullet$ &  &  &  &  & $\bullet$ &  &  &  & $\bullet$ &  & $\bullet$ &  &  &  &  &  & $\bullet$ &  &  \\
\rowcolor[HTML]{C0C0C0} 
2020 & Sanchez et al.~\cite{sanchez_recommendation_2020} & $\bullet$ &  &  &  & $\bullet$ &  &  &  &  &  & $\bullet$ &  &  &  &  &  &  & $\bullet$ &  &  \\
2021 & Kaur et al.~\cite{barolli_reinforcement_2021} & $\bullet$ &  &  &  &  &  &  & $\bullet$ &  & $\bullet$ &  & $\bullet$ &  &  &  & $\bullet$ &  & $\bullet$ &  &  \\
\rowcolor[HTML]{C0C0C0} 
2021 & Lobner et al.~\cite{lobner_explainable_2021} &  & $\bullet$ &  & T &  &  &  &  &  &  &  & $\bullet$ & $\bullet$ &  & $\bullet$ &  & $\bullet$ &  &  &  \\
2022 & Filipczuk et al.~\cite{filipczuk_automated_2022} & $\bullet$ &  &  &  &  &  & $\bullet$ &  &  &  & $\bullet$ &  & $\bullet$ &  &  & $\bullet$ &  &  &  &  \\
\rowcolor[HTML]{C0C0C0} 
2022 & Hirschprung et al.~\cite{hirschprung_game_2022} &  &  & $\bullet$ &  &  &  & $\bullet$ &  &  & $\bullet$ &  &  &  &  &  &  & $\bullet$ &  &  &  \\
2022 & Kökciyan et al.~\cite{kokciyan_taking_2022} &  & $\bullet$ &  &  &  &  & $\bullet$ &  &  & $\bullet$ &  &  &  &  & $\bullet$ &  &  & $\bullet$ &  &  \\
\rowcolor[HTML]{C0C0C0} 
2022 & Ulusoy et al.~\cite{ulusoy_panola_2022} &  &  & $\bullet$ &  &  &  &  & $\bullet$ &  & $\bullet$ &  &  &  &  & $\bullet$ &  & $\bullet$ &  &  &  \\
2022 & Zhan et al.~\cite{zhan_model_2022} & $\bullet$ &  &  &  &  &  & $\bullet$ & $\bullet$ &  &  & $\bullet$ &  &  &  &  &  &  & $\bullet$ &  &  \\
\rowcolor[HTML]{C0C0C0} 
2022 & Brandão et al.~\cite{brandao_prediction_2022} &  & $\bullet$ &  & NIT & $\bullet$ &  &  &  &  & $\bullet$ &  &  &  &  &  & $\bullet$ &  &  &  &  \\
2022 & Mendes et al.~\cite{mendes_enhancing_2022} & $\bullet$ &  &  & NIT &  &  &  &  &  & $\bullet$ &  &  &  &  &  & $\bullet$ &  &  &  &  \\
\rowcolor[HTML]{C0C0C0} 
2022 & Shanmugarasa et al.~\cite{shanmugarasa_automated_2022} & $\bullet$ &  &  &  & $\bullet$ &  &  &  &  & $\bullet$ & $\bullet$ & $\bullet$ & $\bullet$ &  &  &  &  & $\bullet$ &  &  \\
2023 & Ayci et al.~\cite{ayci_uncertainty-aware_2023} &  &  & $\bullet$ & NIT &  &  &  &  &  &  & $\bullet$ & $\bullet$ &  &  &  &  & $\bullet$ &  &  &  \\
\rowcolor[HTML]{C0C0C0} 
2023 & Serramia et al.~\cite{serramia_predicting_2023} &  & $\bullet$ &  &  &  & $\bullet$ &  &  &  &  & $\bullet$ & $\bullet$ &  &  &  &  &  & $\bullet$ &  &  \\
2024 & Wang et al.~\cite{wang_privacyoracle_2024} & $\bullet$ &  &  &  &  &  &  &  & $\bullet$ & $\bullet$ & $\bullet$ & $\bullet$ & $\bullet$ & $\bullet$ &  &  &  & $\bullet$ &  &  \\ \hline
\end{tabular}
\end{table*}

\begin{table*}[]
\scriptsize
\caption{Summary table of our classification, part 2.
~
It presents the optional features of \PPAs, such the \textbf{Architecture}, under which we denote with ``--'' when the criterion is not applicable (no implementation/tool is presented) and when the solution presents an implementation, but the paper did not specify enough information to infer its architecture.
~
For \textbf{user control over decisions}, we specify the elements present to inform users under \textit{Informed} (type of \textbf{D}ata, \textbf{P}urpose, \textbf{C}ontroller).
~
It also presents the \textbf{accuracy} (if any) of the predictions (see Section~\ref{subsec:taxo_validation}); the type of \textbf{user control over decisions}; the presence or not of a \textbf{user study}, and the type of user study if applicable; and the results of our \textbf{critical appraisal} (see Section~\ref{subsec:appraisal}).}
\label{tab:classification_2}
\hskip-1.15cm
\begin{threeparttable}[b]
\begin{tabular}{l|l|lll|llll|l|l|l}
\hline
 &  & \multicolumn{3}{l|}{Architecture} & \multicolumn{4}{l|}{User control over decision} &  &  &  \\ \cline{3-9}
 &  &  &  &  &  &  &  &  &  &  &  \\
 &  &  &  &  &  &  &  &  &  &  &  \\
 &  &  &  &  &  &  &  &  &  &  &  \\
 &  &  &  &  &  &  &  &  &  &  &  \\
 &  &  &  &  &  &  &  &  &  &  &  \\
 &  &  &  &  &  &  &  &  &  &  &  \\
\multirow{-8}{*}{Year} & \multirow{-8}{*}{Publication} & \cellcolor[HTML]{FFFFFF}\begin{rotate}{90} Local \end{rotate} & \cellcolor[HTML]{FFFFFF}\begin{rotate}{90} Remote \end{rotate} & \cellcolor[HTML]{FFFFFF}\begin{rotate}{90} Federated \end{rotate} & \cellcolor[HTML]{FFFFFF}\begin{rotate}{90} Informed \end{rotate} & \cellcolor[HTML]{FFFFFF}\begin{rotate}{90} Semi-automated \end{rotate} & \cellcolor[HTML]{FFFFFF}\begin{rotate}{90} Specific \end{rotate} & \cellcolor[HTML]{FFFFFF}\begin{rotate}{90} Revoke \end{rotate} & \multirow{-8}{*}{Accuracy} & \multirow{-8}{*}{User study} & \multirow{-8}{*}{Critical appraisal} \\ \hline
2014 & Xie et al.~\cite{xie_location_2014} & -- & -- & -- & No & Yes & Yes~\tnote{a} & No & 68\% & Online user experiment~$\alpha$ & -- \\
\rowcolor[HTML]{C0C0C0} 
2015 & Apolinarski et al.~\cite{apolinarski_automating_2015} & $\bullet$ &  &  & D & Yes & Yes & No & -- & No & -- \\
2015 & Hirschprung et al.~\cite{hirschprung_simplifying_2015} & -- & -- & -- & D & No~\tnote{b} & Yes & No & -- & Online qualitative survey & D-, very low (55\%) \\
\rowcolor[HTML]{C0C0C0} 
2015 & Squicciarini et al.~\cite{squicciarini_privacy_2015} & -- & -- & -- & D & Yes & Yes & No & 92.53\% & Cross sectional study~\tnote{c} & D-, very low (55\%) \\
2016 & Liu et al.~\cite{liu_follow_2016} & ? & ? &  & D, P & Yes & Yes & Yes & 78.7\% & Randomized controlled studies~\tnote{d} & A, high (90\%) \\
\rowcolor[HTML]{C0C0C0} 
2016 & Albertini et al.~\cite{albertini_privacy_2016} &  & $\bullet$ &  & D & Yes & No & No & -- & Cross-sectional study & D, very low (55\%) \\
2016 & Dong et al.~\cite{dong_ppm_2016} & -- & -- & -- & -- & -- & -- & -- & 89,8\% F1 & Case studies~$\alpha$ & -- \\
\rowcolor[HTML]{C0C0C0} 
2017 & Baarslag et al.~\cite{baarslag_automated_2017} & $\bullet$ & $\bullet$ &  & Unclear & Yes & Yes & No & -- & Randomized controlled study~\tnote{e} & A, high (90\%) \\
2017 & Fogues et al.~\cite{fogues_sosharp_2017} &  & $\bullet$ &  & No & No & No & No & Around 50\% & Online survey~$\alpha$ & -- \\
\rowcolor[HTML]{C0C0C0} 
2017 & Zhong et al.~\cite{zhong_group-based_2017} & -- & -- & -- & -- & -- & -- & -- & 79\% & Survey~$\alpha$ & -- \\
2017 & Misra et al.~\cite{misra_pacman_2017} &  & $\bullet$ &  & D & Yes & Yes & No & 91.8\% & Non-controlled before-after study~\tnote{f} & C, limited (70\%) \\
\rowcolor[HTML]{C0C0C0} 
2017 & Nakamura et al.~\cite{camp_easing_2017} & -- & -- & -- & -- & -- & -- & -- & 85\% & Cross-sectional study~\tnote{g} & D, very low (55\%) \\
2017 & Olejnik et al.~\cite{olejnik_smarper_2017} & $\bullet$ &  &  & No & Yes & Yes & No & More than 80\% & Yes, for data collection~$\alpha$ & -- \\
\rowcolor[HTML]{C0C0C0} 
2018 & Das et al.\cite{das_personalized_2018} &  & $\bullet$ &  & Yes & It depends & Yes & No & -- & No & -- \\
2018 & Tan et al.~\cite{tan_context-perceptual_2018} &  & $\bullet$ &  & No & No~\tnote{h} & Yes & No & 95\%~\tnote{i} & No & -- \\
\rowcolor[HTML]{C0C0C0} 
2018 & Wijesekera et al.~\cite{wijesekera_contextualizing_2018} & $\bullet$ & $\bullet$ &  & D, C & Yes & Yes & Yes & 95\% & Interrupted time series study (ESM) & B, moderate (80\%) \\
2018 & Yu et al.~\cite{yu_leveraging_2018} & -- & -- & -- & -- & -- & -- & -- & -- & Cross-sectional study~\tnote{j} & D, very low (55\%) \\
\rowcolor[HTML]{C0C0C0} 
2018 & Bahirat et al.~\cite{bahirat_data-driven_2018} & -- & -- & -- & D, P~\tnote{k} & It depends & It depends & No & 81.54\% & No & -- \\
2018 & Raber et al.~\cite{raber_retailio_2018} & -- & -- & -- & D & Yes\tnote{l} & Yes & Yes & 70\% & Non-controlled before-after study~\tnote{m} & C, limited (70\%) \\
\rowcolor[HTML]{C0C0C0} 
2019 & Klingensmith et al.~\cite{klingensmith_hypervisor-based_2019} & $\bullet$ & $\bullet$ &  & D & Not always & Yes & No & -- & No & -- \\
2019 & Barbosa et al.~\cite{barbosa_what_2019} & -- & -- & -- & -- & -- & -- & -- & 86.8\%~\tnote{n} & Survey~$\alpha$ & -- \\
\rowcolor[HTML]{C0C0C0} 
2019 & Alom et al.~\cite{alom_helping_2019} & -- & -- & -- & -- & -- & -- & -- & Up to 72.2\% (satisfaction) & Cross-sectional study~\tnote{o} & D, very low (55\%) \\
2019 & Alom et al.~\cite{alom_adapting_2019} & -- & -- & -- & -- & -- & -- & -- & 96.4\% and 94.5\%~\tnote{p} & Yes, for labeling and evaluation~$\alpha$ & -- \\
\rowcolor[HTML]{C0C0C0} 
2020 & Kasaraneni et al.~\cite{barolli_selflearning_2020} &  & $\bullet$ &  & D & Yes & Yes & No & -- & No & -- \\
2020 & Kaur et al.~\cite{kaur_smart_2020} & -- & -- & -- & -- & -- & -- & -- & -- & No & -- \\
\rowcolor[HTML]{C0C0C0} 
2020 & Botti-Cebria et al.~\cite{herrero_automatic_2021} &  & $\bullet$ &  & D & Yes & Yes & No & --~\tnote{q} & No & -- \\
2020 & Kökciyan et al.~\cite{kokciyan_turp_2020} & -- & -- & -- & -- & -- & -- & -- & -- & No & -- \\
\rowcolor[HTML]{C0C0C0} 
2020 & Sanchez et al.~\cite{sanchez_recommendation_2020} & -- & -- & -- & Unclear & Yes & Yes & No & 84.74\% & Online survey to build their dataset~$\alpha$ & -- \\
2021 & Kaur et al.~\cite{barolli_reinforcement_2021} & -- & -- & -- & -- & -- & -- & -- & -- & No & -- \\
\rowcolor[HTML]{C0C0C0} 
2021 & Lobner et al.~\cite{lobner_explainable_2021} & -- & -- & -- & -- & -- & -- & -- & 83.33\%~\tnote{r} & Survey~$\alpha$ & -- \\
2022 & Filipczuk et al.~\cite{filipczuk_automated_2022} & $\bullet$ & $\bullet$ &  & D & Yes & Yes & No & 65\%~\tnote{s} & Non-controlled before-after study~\tnote{t} & C, limited (70\%) \\
\rowcolor[HTML]{C0C0C0} 
2022 & Hirschprung et al.~\cite{hirschprung_game_2022} & -- & -- & -- & -- & -- & -- & -- & -- & Cross-sectional study & D, low (60\%) \\
2022 & Kökciyan et al.~\cite{kokciyan_taking_2022} & -- & -- & -- & No & It depends & Yes & No & Between 41 and 92\%~\tnote{u} & No & -- \\
\rowcolor[HTML]{C0C0C0} 
2022 & Ulusoy et al.~\cite{ulusoy_panola_2022} & -- & -- & -- & -- & -- & -- & -- & Around 75\%~\tnote{v} & No & -- \\
2022 & Zhan et al.~\cite{zhan_model_2022} & -- & -- & -- & -- & -- & -- & -- & 74\% & No & -- \\
\rowcolor[HTML]{C0C0C0} 
2022 & Brandão et al.~\cite{brandao_prediction_2022} &  &  & $\bullet$ & -- & -- & -- & -- & Between 82 and 88\% & Field study~$\alpha$ & -- \\
2022 & Mendes et al.~\cite{mendes_enhancing_2022} &  & $\bullet$ &  & No & No & No & No & 92\% & Field study~$\alpha$ & -- \\
\rowcolor[HTML]{C0C0C0} 
2022 & Shanmugarasa et al.~\cite{shanmugarasa_automated_2022} & $\bullet$ &  &  & No & Yes & Yes & No & 92.62\% & Cross-sectional study~\tnote{w} & D, very low (55\%) \\
2023 & Ayci et al.~\cite{ayci_uncertainty-aware_2023} &  & $\bullet$ &  & No & Yes & Yes & No & 89\% & No & -- \\
\rowcolor[HTML]{C0C0C0} 
2023 & Serramia et al.~\cite{serramia_predicting_2023} &  & $\bullet$ &  & No & Yes & No & No & 3.78/5~\tnote{x} & Cross-sectional study~\tnote{y} & D, very low (55\%) \\
2024 & Wang et al.~\cite{wang_privacyoracle_2024} &  & $\bullet$ &  & No & No & Yes & No & -- & No & -- \\ \hline
\end{tabular}
\begin{tablenotes}
\begin{multicols}{2}
\item[a] Only location
\item[b] Not necessarily, depends on what they call the Configuration Options
\item[c] A survey-based study and a direct user evaluation
\item[d] Two surveys
\item[e] Between-participants design
\item[f] Online survey
\item[g] Online questionnaire
\item[h] Not by default, they have a sort of ‘user settings’ for expert users
\item[i] For privacy leakage detection (not to be confused with preferences detection)
\item[j] To measure the interpretability of the approaches
\item[k] Not consistently
\item[l] Based on the current data collection
\item[m] Online survey
\item[n] AUC of binary allow/deny for a given scenario
\item[o] User satisfaction
\item[p] Accuracy based on appreciation of evaluators
\item[q] The accuracy presented is for the right category of data
\item[r] With interpretability of the results
\item[s] On average, but seems higher in specific case
\item[t] Between-subject experimental design
\item[u] Depends on several parameters
\item[v] Difficult to assess because they measure utility of decisions in a simulated setting
\item[w] Online survey
\item[x] Acceptability rate, not accuracy
\item[y] To measure the level of comfort of the norms inferred
\item[$\alpha$] Alpha means that the user study is not meant to assess the solution, but only meant to collect data
\end{multicols}
\end{tablenotes}
\end{threeparttable}
\end{table*}

The classification and its dimensions are \textbf{data-driven}, in the sense that they were derived based on what is described in the papers, reflecting the current state of the literature.
For example, considering the category of system contexts, more dimensions could be envisioned, but we limited it to the five dimensions (i.e., mobile apps, social media, IoT, cloud, and intelligent retail stores) that were found in the papers.
Each feature will be explored in more detail in this section, and substantiated with non-exhaustive examples for each possible option, while an overview is provided in Figure~\ref{fig:classification}.

Note that not all dimensions are necessary for composing an \PPA.
The dimensions for the type of AI, source of data, type of decision, and system context are \textit{``mandatory,''} consisting of essential requisites that an \PPA needs to consider (solid boxes in Figure~\ref{fig:classification}).
Other dimensions such as the empirical assessment, choice architecture, and user control over decisions are \textit{``optional''} since not all the identified \PPAs were evaluated, some do not have an implementation (and therefore an architecture), and some (regrettably) do not empower users with much control for various reasons (dashed boxes in Figure~\ref{fig:classification}).

Furthermore, note that most but not all dimensions are non-exclusive.
For instance, it is possible to combine different types of data and/or AI models (non-exclusive), but the system context is often exclusive in the sense that solutions are often designed for a specific system context.

% the papers address each of these dimensions to different extents, and their options are often non-exclusive.
% For instance, all articles surveyed discuss the type of decision, but this is not the case for the choice architecture; and while most solutions are composed of different sources of data and combine different AI models, the system context is often exclusive in the sense that solutions are often designed for a specific system context.

\subsection{Type of Decision}
\label{subsec:taxo_decision}
Decisions taken by an \PPA can be of different types, and it is essential to distinguish them to assess the possibilities they offer.
Indeed, some decisions -- such as permissions -- have a binding character, i.e., they constrain the system to act according to the user's choice, while others do not, such as preferences.
Note that it may not always be possible to distinguish between each type of decision clearly (as discussed in Section~\ref{subsec:further-decisions}).
Other types of decisions with different implications regarding their enforcement can be envisioned by an \PPA (such as consent or deletion requests, see Section~\ref{subsec:privacy_decisions}).

\subsubsection{Permissions}
\label{subsub:permissions}
The first type of decisions that many \PPAs assist the users with is \textit{permissions}, which, as discussed in Section~\ref{subsec:further-decisions}, correspond to access control settings.
Permissions are system-specific and binding, as the underlying operating system should enforce them.

We typically find mobile app permissions (e.g., in Baarslag et al.\cite{baarslag_automated_2017}, mobile apps are addressed in 11 papers), but they are not restricted to the mobile environment.
\PPAs can deal with permissions in IoT environments (see, e.g., ~\cite{klingensmith_hypervisor-based_2019}, IoT is covered by 13 papers) or in the cloud~\cite{hirschprung_simplifying_2015}.

\subsubsection{Preferences}
The second type of decision covered by the literature is \textit{preferences}, which, unlike %their binding counterparts
permission settings, should be understood as expressions of will.
Several works refer to preferences while they actually deal with permissions~\cite{liu_follow_2016,filipczuk_automated_2022,wijesekera_contextualizing_2018,hirschprung_simplifying_2015,shanmugarasa_automated_2022}.
It is indeed common to talk about preferences imprecisely, but they should not be confused with permissions that have a binding property.

\subsubsection{Data Sharing}
\textit{Data sharing} is the third type of privacy decision of \PPAs encountered in the reviewed literature, for which the binding character is uncertain for users. 
For instance, assessing whether a limitation in the audience is enforced is not always possible from a user point of view because the underlying technical system is inaccessible to them, see, e.g., Ulusoy et al.~\cite{ulusoy_panola_2022}.
Typically, it can be difficult or even impossible to assess whether most social media platforms strictly account for the user's privacy decisions, or merely welcome them as recommendations to be applied only if possible.
Papers classified under this type of decision usually do not mention the binding character of their solution (or the lack thereof).

\subsection{AI Technology Used}
\label{subsec:taxo_ai}
% \todo[inline]{Clarify the analysis @victor for the AI analysis}

Another significant characteristic of \PPAs is the type of AI used.
Many solutions are based on machine learning models, such as supervised ML (classification), non-supervised ML (clustering), and reinforcement learning, sometimes combined.
It is, however, also possible to find older AI techniques grouped under the umbrella of expert or rule-based systems.

We also classified the different AI technologies used by the \PPAs reviewed regarding their explainability, or their inherent transparency.
However, XAI is only explicitly addressed by one work~\cite{lobner_explainable_2021}; the other models are therefore categorized based on Arrieta et al.~\cite{arrieta_explainable_2019}'s taxonomy, which defines non-ML based systems as AI (including rule-based).
For classification models, we annotated T for Transparent in Table~\ref{tab:classification_1}, NIT for Not-Inherently Transparent, and PT for Partially Transparent when the solution relies on models with different levels of transparency.
Note that LLMs were not considered in Arrieta et al., we nonetheless classify them as non-inherently transparent.

\subsubsection{Transparent}
\paragraph{Classification}
Supervised machine learning, also called classification models, is a common set of techniques deployed in \PPAs.
In this context, a model is trained to classify an object of decision into a choice tailored to the users' desires.

Transparent classification models~\cite{arrieta_explainable_2019} (used in 8 papers) are composed of decision trees (used for instance in Bahirat et al.~\cite{bahirat_data-driven_2018}), k-nearest neighbors (leveraged in Botti-Cebria et al.~\cite{herrero_automatic_2021}), and Bayesian models (see Olejnik et al.~\cite{olejnik_smarper_2017}).

\paragraph{Clustering}
\label{subsub:clustering}
Several works use clustering techniques for their \PPA.
In this context, clustering is classically used to create a set of \textit{privacy profiles}, i.e., an archetypal ensemble of default parameters (for preferences or permissions) to which a user is then assigned.
Clustering algorithms (leveraged in 6 papers) used are hierarchical clustering~\cite{liu_follow_2016}, k-means~\cite{brandao_prediction_2022}, k-modes~\cite{shanmugarasa_automated_2022}, although several papers did not disclose the exact method used, such as Hirschprung et al.~\cite{hirschprung_simplifying_2015}.

\paragraph{Rule-based}
\PPAs can be powered by non-machine-learning algorithms, based instead on rules (e.g., Albertini et al.~\cite{albertini_privacy_2016} implement association rules).
This comprises theoretical as well as practical works, with two out of four providing a tool (\cite{albertini_privacy_2016} and \cite{serramia_predicting_2023}).

\subsubsection{Not-inherently Transparent}
\paragraph{Classification}
Non-transparent classification models (found in 14 papers) typically encompass classic neural networks (as in Klingensmith et al.~\cite{klingensmith_hypervisor-based_2019}) and deep neural networks  (see for instance Yu et al.~\cite{yu_leveraging_2018}); random forests~\cite{misra_pacman_2017}, Ada Boost~\cite{mendes_enhancing_2022} and Support Vector Machines (used in Wijesekera et al.~\cite{wijesekera_contextualizing_2018}) complete the picture.
Post-hoc explanations must complement these models, as they are not easily understandable by themselves.

\paragraph{Reinforcement}
Reinforcement learning is the least used family of machine-learning techniques in \PPAs.
It is implemented in Kaur et al.~\cite{barolli_reinforcement_2021} and Ulusoy et al.~\cite{ulusoy_panola_2022}, both used to adapt users' feedback to their preferences, and in Zhan et al.~\cite{zhan_model_2022}.
The first paper uses it to disclose information (using permissions), while the second uses it to learn bidding preferences in a negotiation context.

\paragraph{Logic-based}
\PPAs can be based on logic (5 papers), for instance, expert systems (Kökciyan et al.~\cite{kokciyan_taking_2022} uses an agent-based model) or game theory (such as Hirschprung et al.~\cite{hirschprung_game_2022}).
These works, albeit few, span various system contexts and types of decisions.

\paragraph{LLM}
We selected only one paper using Large-Language Models (LLMs)~\cite{wang_privacyoracle_2024} during an update of the survey.
Here, the authors leverage GPT-3.5 and GPT-4.0 to produce \textit{privacy rules}, based on sensors outputs and user preferences.
Note that LLMs are based on deep neural networks and are, therefore, difficult to explain and even prone to hallucinations.

\subsection{Source of Data}
\label{subsec:taxo_source}
An \PPA can rely on various \textit{sources of data} when using AI to help with a privacy decision.
These data sources are very often combined, and a careful choice is necessary to fully exploit the potential of the models described in the previous section.

\subsubsection{Context}
\label{subsubsec:context}
\textit{Context} is an often-used data source, yet not always well-defined.
However, when it is defined, it is composed of the location~\cite{xie_location_2014}, the time, relationships with other individuals~\cite{fogues_sosharp_2017}, or the activity performed~\cite{alom_helping_2019}.

External data provided by third parties or other unrelated entities is sometimes used to predict privacy decisions, and this external data can arguably be considered context.
For instance, under this term, we find risk factors~\cite{ayci_uncertainty-aware_2023} or information related to other applications in the background~\cite{brandao_prediction_2022}.

Context is usually a crucial component for an effective \PPA because, as has been argued under Nissembaum's theory of privacy as contextual integrity~\cite{nissenbaum_privacy_2004}, context is paramount to designing appropriate information flows and respecting privacy norms.

\subsubsection{Attitudinal Data}
A few \PPAs ask users questions to elicit so-called \textit{attitudinal data} about stated practices, or preferences regarding privacy recommendations to avoid the so-called cold-start problem, which arises when no past data is available to provide a recommendation.
For example, Nakamura et al.~\cite{camp_easing_2017} focuses on asking a minimal set of questions while keeping accuracy as high as possible, or Alom et al.~\cite{alom_adapting_2019} asks \textit{``a reasonable [sic] number of questions (50) to the users.''}

\subsubsection{Behavioral Data}
Another common source of data is \textit{behavioral data}.
Behavioral data has the advantage of reflecting the \textit{actual} 
users' privacy decisions to predict the next ones, as it does not simply rely on stated practices (unlike attitudinal data).
While it can be a powerful tool, it can also create a feedback loop, reinforcing the same decisions. 

Behavioral data can encompass past decisions, such as in Zhan et al.~\cite{zhan_model_2022}, which leverage past choices to fill a knowledge base and then use them to predict privacy decisions.
It can also comprise current settings or preferences
on a specific type of data to infer a decision for another type~\cite{hirschprung_simplifying_2015}.
The system can also use these preferences to match users to a particular privacy profile, using for instance clustering techniques (see Section~\ref{subsub:clustering}).

\subsubsection{Metadata}
\textit{Metadata} is data that provides information about other data, for example, the name of an application used~\cite{wijesekera_contextualizing_2018}, network requests~\cite{tan_context-perceptual_2018}, the purpose associated with processing~\cite{barbosa_what_2019}, the usage frequency of certain permissions (such as location) by an app~\cite{kaur_smart_2020}, or tags associated with images~\cite{squicciarini_privacy_2015}.
To some extent, metadata can overlap with context, for instance, when considering time or location. 
However, the articles surveyed more often refer to the time and location of collection \textit{of a certain data point} for metadata, and to the \textit{current time and location} when a decision has to be made for context.
Metadata can provide peripheral information to make decisions, although it is rarely used as a sole source of data (out of 14 leveraging metadata, only 4 papers~\cite{tan_context-perceptual_2018,klingensmith_hypervisor-based_2019,barolli_selflearning_2020} rely solely on it).

\subsubsection{Data Type}
The \textit{data type} refers to the category of data concerned by the decision, such as whether it is an image to share on social media~\cite{zhong_group-based_2017}, the location requested by an app~\cite{filipczuk_automated_2022}, or various sensor data by an IoT device~\cite{shanmugarasa_automated_2022}.
The type of data can provide accurate information about the sensitiveness of a decision (location data can, for instance, provide sensitive information regarding the users' context, e.g., from location data that reveals that a user visits a clinic or church,  medical, or religious information could be inferred),
%be seen as more sensitive than professional emails, and so regardless of its content)
yet only a relatively low number of solutions rely on the data type to build an \PPA~\cite{baarslag_automated_2017,zhong_group-based_2017,camp_easing_2017,wijesekera_contextualizing_2018,lobner_explainable_2021,filipczuk_automated_2022,shanmugarasa_automated_2022}.

\subsubsection{Content of Data}
The \textit{content of data} refers to the specific content of a data point, as the name indicates. However, we also include data that can be directly inferred from the content of data under this category.
For example, Botti-Cebria et al.~\cite{herrero_automatic_2021} and Dong et al.~\cite{dong_ppm_2016} estimate the sensitivity of the content of the information to be shared to help make a decision.
Indeed, content can be leveraged to tailor decisions: a picture deemed private should not receive the same treatment as one deemed public, and a geolocation trace that may potentially allow inferring religious practice should cautiously be dealt with.

\subsection{System Context}
\label{subsec:taxo_system_context}
Most \PPAs target a specific \textit{system context}, that is, a set of technologies with distinct characteristics.
Indeed, each system context has specific requirements that one must consider when designing an \PPA.
System contexts differ by the availability of an \textbf{interface}, \textbf{computational power}, and control over the \textbf{architecture}.

\subsubsection{Mobile Apps}
Several works focus on mobile applications, and often on Android~\cite{apolinarski_automating_2015,wijesekera_contextualizing_2018}.
Mobile ecosystems have the advantage of being well-defined ecosystems, enabling the possibility to strictly enforce privacy decisions (i.e., it is often addressed with permissions, see Section~\ref{subsub:permissions}).

Mobile phones also possess reasonable computational power (in the sense that they can run an \PPA) and a screen enabling direct user interactions.
Hence, an \PPA can be implemented directly on a smartphone (see Baarslag et al.~\cite{baarslag_automated_2017}), and it can interact with and even regulate mobile apps, all of which make mobile ecosystems suitable candidates for \PPAs under the users' control.

\subsubsection{Internet of Things}
Another widely used system context for \PPAs is the Internet of Things (IoT).
We understand IoT as a network of devices, including sensors, mechanical and digital machines, as well as consumer devices, all connected to the Internet. 
In practice, \PPAs have been developed for smart homes~\cite{shanmugarasa_automated_2022, barbosa_what_2019}, on campuses~\cite{das_personalized_2018}, or for wearables such as fitness devices~\cite{sanchez_recommendation_2020} for instance.

Most IoT devices are usually not equipped with proper interfaces and lack computational power.
These characteristics make it challenging to build \PPAs %enforcing 
assisting with permission settings, yet not impossible (see Klingensmith et al.~\cite{klingensmith_hypervisor-based_2019} for instance, who manage to do so with an \PPA located on end devices).

\subsubsection{Social Media}
According to our classification of the literature, the third major system context is social media, for which several \PPAs have been designed to help make privacy decisions.
In this case, neither the interface nor the computational power are usually limiting factors.
However, the design and implementation of social media platforms (that are usually not published openly) make it difficult to assess the binding character of privacy decisions supported by \PPAs running on social media platforms.
\PPA solutions are rather designed to support data sharing, i.e., whether a specific post should be shared on social media and with whom, than focusing on assisting users with privacy decision-making.

\subsubsection{Cloud}
A less prevalent system context is cloud environments, with only one of the reviewed articles proposing an \PPA targeting cloud environments~\cite{hirschprung_simplifying_2015}.
% , even though only one of the reviewed articles proposes an \PPA for the cloud.
Their solution offers a method to simplify information disclosure in cloud environments such as Google Drive.
However, this work is thus a lone example and contrasts with the otherwise balanced distribution of works among other system contexts.

\subsubsection{Intelligent retail store}
Similar to cloud environments, only one work, Raber et al.~\cite{raber_retailio_2018}, was captured under this category.
Their \PPA provides a solution to automate decisions in intelligent retail stores, combining pervasive computing and online applications.

\subsection{Architecture}
\label{subsec:taxo_architecture}
By architecture, we refer here to where the computation happens, i.e., the decision-making, and not necessarily the pre-processing steps such as building privacy profiles.
Directly connected to the architecture is the trust model of the \PPA.
While this term is usually reserved for security-oriented research, describing whether one has to trust the different entities or not provide relevant information for understanding the privacy boundaries.

Note that the location of the computation is only relevant for implemented \PPAs, and not for theoretical models. 
Similarly, most solutions surveyed do not explicitly describe a threat or trust model in their paper. 
Nonetheless, it is possible to infer that trusted parties are required in some solutions.
For instance, Tan et al.~\cite{tan_context-perceptual_2018} describe an architecture comprising a remote classifier (in which one has to place trust), yet no trust model is described.

\subsubsection{Local Computation}
The processing can happen locally on the user device, such as on a smartphone (see, e.g., Olejnik et al.~\cite{olejnik_smarper_2017}), but this device can also be a home pod in an IoT context (see, e.g., Shanmugarasa et al.~\cite{shanmugarasa_automated_2022}). 

Creating and processing user profiles, using local AI models, and locally deriving privacy decisions have the advantage that the user can keep control over the locally processed data, including their profiles and AI models, which usually can include sensitive information about the user's preferences or behavior. 
However, local data processing also puts more responsibilities on the user to secure the devices properly against malware or other attacks.

\subsubsection{Remote Computation}
The \PPA could also be based on remote data processing (according to the user's point of view), involving a central server that processes personal privacy decisions and contextual data, including, e.g., location data or another type of data. 
Remote computation raises the question of the trust placed in the party performing this computation to protect the data properly, to enforce the data subject's rights (e.g., to access or to delete their data and computed profiles or models), and not to use the data for any unintended purposes~\cite{tan_context-perceptual_2018}.

Several solutions rely on a remote third party that has to be trusted, e.g., Baarslag et al.~\cite{baarslag_automated_2017}, or Tan et al.~\cite{tan_context-perceptual_2018}'s solution that places trust on their own remote classifier. 
The solution developed by Wang et al.~\cite{wang_privacyoracle_2024} relies on ChatGPT, a closed-source chatbot on which all trust has to be placed with little or no accountability.
In contrast, others only require trusting the operating system (OS) on which the \PPA is implemented~\cite{olejnik_smarper_2017}, or require trusting both the OS and mobile applications~\cite{apolinarski_automating_2015}.

\subsubsection{Federated Learning}
Only one article, by Brand\~{a}o et al.~\cite{brandao_prediction_2022}, presented an \PPA based on federated learning.
In this work, the processing of user data for the computation of locally trained neural network models happens on the user devices.
These devices only share the neural network weights with a central server, which will, in turn, average all the local weights and send back the results to the clients, which can use these new weights to continue the training process.
Federated learning is a privacy-enhancing approach for processing the users' raw data only locally, which can achieve a performance comparable to the centralized approach (remote computation). 
Nonetheless, federated learning could still be attacked, e.g., with membership inference attacks, to leak personal data from locally trained models~\cite{shokri2017membership}.

\subsection{Empirical Assessment}
\label{subsec:taxo_validation}
\PPAs' performance can be measured in terms of accuracy, but because several solutions are meant to be usable tools, assessing an \PPA encompasses more than a mere measurement of how well a privacy decision is predicted.

As mentioned in Section~\ref{subsec:analysis}, an empirical assessment can be an evaluation (see, e.g.,~\cite{liu_follow_2016}) or a validation (e.g.,~\cite{hirschprung_game_2022}). 

\subsubsection{User Study}
A classical way to validate a tool or a method is to conduct a user study, and we found 16 papers reporting a user study to validate usability.
A user study can have various interpretations, ranging from a simple questionnaire to rate satisfaction (such as Alom et al.~\cite{alom_helping_2019}) to a large-scale randomized controlled study (see, e.g., Liu et al.~\cite{liu_follow_2016}) -- the former being more akin to a mere validation, the latter a full-fledged evaluation.

Note that several works elicited data to build a dataset through a user study, which was therefore not meant as a means of assessment (annotated as $\alpha$ ~in Table~\ref{tab:classification_2}).

\subsubsection{Purely Statistical (Dataset)}
Several works provide a validation without a user study, that is, only based on a purely statistical analysis based on a dataset~\cite{barolli_selflearning_2020}, \cite{herrero_automatic_2021}, \cite{bahirat_data-driven_2018}, \cite{kokciyan_turp_2020}, \cite{ayci_uncertainty-aware_2023}, \cite{zhan_model_2022}, \cite{kokciyan_taking_2022}, \cite{herrero_automatic_2021}.
Such a measure, although potentially subject to a higher degree of statistical rigor, cannot necessarily capture users' expectations and may even fall into the pitfall of Goodhart's law.\footnote{According to which ``When a measure becomes a target, it ceases to be a good measure.''}

\subsubsection{Accuracy Measurement}
Accuracy can measure the capacity of an \PPA to predict a privacy decision, but not all papers measure the same type of accuracy.
Tan et al.~\cite{tan_context-perceptual_2018} measure privacy leakage detection, Botti-Cebria et al.~\cite{herrero_automatic_2021} whether the correct category of data is predicted or not, Serramia et al.~\cite{serramia_predicting_2023} the acceptability rate, Barbosa et al.~\cite{barbosa_what_2019} the Area Under the Curve (AUC) of a binary allow/deny for a given scenario, etc.

Other works, while they do measure the accuracy of their solution to predict a privacy decision, present their work with %questionable 
limited rigor or precision.
For example, Fogues et al.,~\cite{fogues_sosharp_2017} only present their results in plots.
In contrast, others, such as Olejnik et al.~\cite{olejnik_smarper_2017}, dedicate an entire subsection to explaining accuracy measurements.

\subsection{User Control Over Decisions}
\label{subsec:taxo_req}

Finally, \PPAs should not only assist users with making privacy decisions but should at the same time also empower users with various options to improve \textit{control over their decisions}.
These options span over \textbf{qualities} of control (solid arrows in Figure~\ref{fig:classification}) and \textbf{instruments} of control (dashed arrows).
The former denotes adjectives that can be appended to control (akin to non-functional requirements in software engineering~\cite{pallas_privacy_2024}), and the latter denotes concrete possibilities or actions for users (similar to functional requirements).

These options are partly related to GDPR requirements for consent (introduced in Section~\ref{subsec:privacy_decisions}), which are thus relevant for privacy decisions that constitute consent.
Note, however, that only a handful of papers specifically refer to legal considerations.
Filipczuk et al.~\cite{filipczuk_automated_2022} refer to the GDPR, Mendes et al.~\cite{mendes_enhancing_2022} acknowledge that an automated response to a permission request might not constitute legal consent, Lobner et al.~\cite{lobner_explainable_2021} base the rationale of explainability on legal requirements, and Sanchez et al.~\cite{sanchez_recommendation_2020} even claim GDPR compliance.
Nonetheless, decisions for setting permissions for mobile operating systems, for instance, still require consent at installation or run time. Thus, legal requirements for consent remain relevant for these types of decisions.  

\subsubsection{Ex-ante Transparency}
Under Art 13 GDPR, data subjects should receive information if data is collected from them, and informing users is also an integral part of the dominant transparency paradigm in the US (the \textit{notice} of the notice and choice approach).
Informing data subjects with intelligible notices arguably improves their control over decisions.
Several \PPAs only inform about the type of data concerned by the privacy decision~\cite{albertini_privacy_2016}, some inform in addition about the controller~\cite{wieringa_requirements_2006} or of the purpose of processing~\cite{bahirat_data-driven_2018}.
In theory, meeting this criterion should not be difficult, although providing intelligible notices requires significant expertise in practice (as illustrated in Schaub et al.~\cite{schaub_design_2015}).

\subsubsection{Semi-automated}

The semi-automated character of a decision refers to including an affirmative action of the user to confirm the decision, which is therefore not fully automated~\cite{morel_automating_2023}. 
Most solutions provide a semi-automated decision process, although not systematically (e.g., Das et al.~\cite{das_personalized_2018} mention that only opt-out is possible for facial recognition), or not always (e.g., Klingensmith et al.~\cite{klingensmith_hypervisor-based_2019} offers different types of ``privacy profiles'', one of which -- \textit{Laissez-Faire} -- enables full automation).
Tan et al.~\cite{tan_context-perceptual_2018} do not leave users in the loop by default, but the system allows the possibility to change the settings for ``experienced users,'' while it depends on the Configuration Option for Hirschprung et al.~\cite{hirschprung_simplifying_2015}.

\subsubsection{Specific}
The specificity of a decision refers to the presentation and the possibility for users to decide on the granularity of each data type, purpose, and controller separately.
For an \PPA, it means having a fine-grained selection process, during which users should not be presented with bundled decisions.
For instance, Shanmugarasa et al.~\cite{shanmugarasa_automated_2022}'s solution works per ``situational context'': who (is requesting data), data type, purpose, and re-sharability (to third parties); while the solution of Xie et al.~\cite{xie_location_2014} only works for one type of data (location), therefore only meeting this option in a restricted sense. 

\subsubsection{Revoke}
Finally, we observed that some \PPAs enable users to withdraw decisions.
Here, rather than denying a decision or a recommendation, revoking operates after a given decision to withdraw it.
This feature has rarely been observed in practice -- at least explicitly -- although the solution of Liu et al.~\cite{liu_follow_2016} allows revoking previously granted decisions.
Revoking previously made decisions, such as sharing data on social media, can be challenging to enforce.
Also, note that certain operating systems -- such as mobile OSes -- will still allow users to revoke their decisions manually, although we stress that this action is performed outside the \PPA.

\section{Discussion}
\label{sec:discussion}
% \todo[inline]{Clarify the recommendations → @simone}

This systematic literature review provides unique insights into how state-of-the-art research has designed \PPAs in recent years.
For instance, IoT became a system context of interest only in 2018, and we observed a similar late adoption trend for reinforcement learning after 2021.\footnote{Some papers may have been published on the topic earlier than in 2013, the year from which we started to include papers in our SLR.}
However, AI techniques have been used in every system context for all types of decisions throughout the years without any apparent pattern.
While this lack of a clear pattern is not the most informative in itself, we ought to look instead at the \textbf{gaps} this survey highlights, the \textbf{challenges} \PPAs raise, then to inform better \textbf{design and development recommendations} based on these analyses.

We acknowledge that our survey of scientific articles reveals primarily gaps in the state-of-the-art research on \PPAs, not gaps in \PPAs that are already used in practice. 
Nonetheless, the best practice recommendations for addressing identified gaps also target developers of \PPAs and may, in these cases, not be appropriate for research projects (as opposed to deployed systems). However, knowledge and awareness of these best practice recommendations may still be helpful for researchers nonetheless.

Based on our main findings, this section provides a detailed discussion organized in seven parts:  the issues of properly \textit{evaluating \PPAs} in Section~\ref{subsec:disc_eval}; \PPAs not sufficiently addressing \textit{Privacy-by-Design} in Section~\ref{subsec:disc_pbd}; the (lack of) \textit{explanations and explainability} in Section~\ref{subsec:disc_xai}; the concerns surrounding \textit{system contexts} in Section~\ref{subsec:disc_contexts}; the relationship with \textit{legal considerations} in Section~\ref{subsec:disc_leg}; the challenges in leveraging different \textit{sources of data} in Section~\ref{subsec:disc_sources}; and finally, potential \textit{research avenues} are introduced in Section~\ref{subsec:future}.

\subsection{Evaluating \PPAs}
\label{subsec:disc_eval}
The problem of the evaluation of \PPAs is two-fold.
First, we observe that \textbf{the evaluations of \PPAs are not based on the same or comparable accuracy metrics or measurements}.
As presented in Section~\ref{subsec:taxo_validation}, accuracy is measured regarding a privacy decision, but also a privacy leakage, acceptability rate, etc.
Second, \textbf{our data shows a lack of user study evaluations}, and our critical appraisal shows a trend toward ``low'' and ``very low'' scores to assess cause and effect.
Only 15 out of 39 papers mentioned that they performed a user study to evaluate their solution\footnote{Some papers include a user study for collecting data, which is not focused on their proposed solution.},\footnote{We verified whether the authors of all submitted papers published follow-up articles, and did not find any.}, but only six studies scored above (or equal to) 70\% based on the CEBMA critical appraisal we performed.
We acknowledge that user studies may go beyond the scope of strictly theoretical papers (e.g., models or frameworks without prototype implementation). 
Yet, we contend that the validation offered by these theoretical papers, often cross-validated on a dataset, is far from being able to reflect reality.
Any proposed \PPAs must be validated and evaluated to substantiate empirical evidence of their value and feasibility.
Without setting unrealistic standards for research, it is still essential that academics and developers strive to put their proposed solutions to the test in real-world settings.\\

\noindent\fbox{
    \begin{minipage}{.9\linewidth}
        \setlength\parskip{1em}
        \textit{\textbf{Recommendation:} Based on the current lack of empirical evidence, we propose that the usability of \PPAs should be evaluated through user studies following high-quality standards for qualitative and quantitative research, and such evaluations should notably encompass the accuracy of the privacy decision taken.}
    \end{minipage}
}

\subsection{Lack of Privacy-by-Design}
\label{subsec:disc_pbd}
Since \PPAs typically analyze the users' attitudinal or behavioral privacy preferences, metadata or content, or other data types for personalized assistance, they need for this purpose to process personal data and user profiles, which could be considered sensitive data.
We identified, however, a gap regarding following a privacy-by-design approach for \PPAs, since hardly any of the papers we surveyed focus on, or mention how, the \PPAs themselves can be designed in a privacy-preserving manner. 
More specifically, among papers describing technical architectures,\footnote{Recall that theoretical papers are excluded from this analysis.} \textbf{only one uses federated learning as a privacy-enhancing approach~\cite{brandao_prediction_2022}, however, without discussing that federated learning is still vulnerable to privacy attacks (see e.g. Mothukuri et al.~\cite{mothukuri2021survey}}. 
Many presented \PPAs require trust in a central server, where the data processing is performed, while data processing on the users' local device may be preferable from a privacy perspective, as it does not require trusting another (central) party.
To this end, Wijesekera et al.~\cite{wijesekera_contextualizing_2018} provides an insightful analysis of the trade-off of having either a local (offline) or a remote computation, concluding that offline learning still performs well (almost 95\% accuracy).
Also note that the privacy threat models are rarely described, making it difficult to evaluate security and privacy assumptions critically.\\

\noindent\fbox{
    \begin{minipage}{.9\linewidth}
        \setlength\parskip{1em}
        \textit{\textbf{Recommendation:} We contend that \PPAs must embrace stronger privacy-by-design principles, including better design strategies but also better integration of Privacy Enhancing Technologies for achieving data minimization, such as federated learning combined with differential privacy, fragmented federated learning~\cite{jebreel2022enhanced} or the use of privacy-preserving data analytics by \PPAs based on multi-party computation, homomorphic or functionally encrypted data (see also~\cite{PAPAYA2019}).}
    \end{minipage}
}

\subsection{Unexplainable AI}
\label{subsec:disc_xai}
Another pitfall identified is the lack of explanations provided by most \PPAs, combined with the lack of explainability/interpretability offered by the AI models used.
\textbf{Only one of the surveyed papers explicitly addresses explainability of the generated decisions}~\cite{lobner_explainable_2021}, and only 8 use transparent models (see Section~\ref{subsec:taxo_ai}) to make predictions.

The growing trend to use deep learning architectures may not facilitate the explainability of decisions, but this challenge is not insurmountable.
It is indeed possible to devise \textit{post hoc} explanations, and to take inspiration from other existing work on usable explanations for AI-made decisions.
Note, however, that inherent transparency can come at the expense of other quality aspects (e.g., accuracy, security, safety, ethical and social considerations~\cite{wang_privacyoracle_2024}) of decisions -- trade-offs must be considered case-by-case.
Deep neural networks tend to outperform their simpler counterparts, although this statement does not seem to generalize to all kinds of decisions, such as decisions made in highly unpredictable settings like social predictions~\cite{narayanan_ai_2024}.

As discussed in Section~\ref{legal_background}, transparency of AI can be a legal requirement in some specific use cases related to \PPAs.
For instance, transparency is required for the data controller according to the GDPR, or for the provider or deployer according to the AI Act, even though this will not apply to most \PPAs and use cases.
In fact, none of the surveyed papers related to high-risk AI applications.
Transparency can in general also foster trust in technology~\cite{crane2006trustguide}.\\

\noindent\fbox{
    \begin{minipage}{.9\linewidth}
        \setlength\parskip{1em}
        \textit{\textbf{Recommendation:} Based on this analysis, we recommend 1) considering the use of inherently explainable AI techniques, such as decision trees, for the classification, whenever this implies that the potential quality loss will be appropriate for the specific use case, or 2) the integration of ad-hoc explanations otherwise, e.g., for neural networks and SVMs.
        %the use of inherently explainable AI models (e.g., decision trees) whenever possible, and, if the increased transparency comes with appropriate quality loss, the implications for the decision-making is clarified (compared to if other AI techniques are used); and otherwise, 
        %and if a potential quality loss will be 
        %in terms of potential acceptable quality loss implications,
        %and if any utility trade-off that may need to be made for the sake of increased transparency will be well acceptable and appropriate for the specific use case context
        %2) integrating \textit{ad-hoc} explanations, e.g., for neural networks and SVMs.
        %Indeed, explainability can also foster trust in technology, improving the uptake of such systems.
       }
    \end{minipage}
}

\subsection{Missing System Context}
\label{subsec:disc_contexts}
Our SLR covered five system contexts: mobile applications, social media, IoT, the cloud, and online retail stores.
We are, however, surprised \textbf{not to find other contexts, such as web browsers or Trigger-Action Platforms (TAPs)}.
The former because cookie notices are notoriously a ``hassle'' for users in modern web experience; we therefore expected to encounter solutions tackling this issue.~\footnote{Recall that Bollinger et al.~\cite{bollinger_automating_2022} does not personalize decisions.}
The latter refers to platforms offering applications for connecting otherwise unconnected devices and services using simple recipes, such as ``Every morning at 7 am, send a Slack message with the first meeting of the day from Google Calendar.'' 
Trigger-action programming has gained a lot of traction in the last years (IFTTT, the most important TAP, boasts over 27 million users~\cite{ifttt_ifttt_nodate}), yet no \PPAs specifically addresses it.
% this environment.

Both these system contexts possess their specific features: many controllers with non-standard interfaces for cookie notices, and numerous actors mediated through a single centralizing entity for TAPs.
They therefore require targeted efforts from designers to offer adequate technological solutions to manage privacy decisions.\\

\noindent\fbox{
    \begin{minipage}{.9\linewidth}
        \setlength\parskip{1em}
        \textit{\textbf{Recommendation:} We encourage researchers and developers of \PPAs to expand their efforts into a broader range of system contexts, also encompassing but not limited to web browsers and TAPs.}
    \end{minipage}
}

\subsection{Few Legal Considerations}
\label{subsec:disc_leg}
As some privacy decisions made by \PPAs have legal privacy implications or issues, legal requirements, e.g., under the GDPR, the AI Act, or other national legislation, should be discussed and considered for the design and use of \PPAs. 
However, a couple of reasons could explain the lack of discussion of legal requirements and implications according to the EU legal framework.
Firstly, the geographic distribution of the solutions surveyed, given that only 13 papers have authors with EU affiliations.
Secondly, the timing of the publications, as 17 papers were published before the GDPR was enacted and none before the AI Act came into force.
%Although partially explainable by 1) the geographic distribution of the solutions surveyed (only 12 papers have EU affiliations, in contrast with both the GDPR and the AI Act being EU regulations), and by 2) the timing of publications (17 papers were published before the GDPR was enacted, and none before the AI Act came into force), 
Nonetheless, it is still surprising to find \textbf{only 4 papers mentioning (but not even addressing) legal considerations}.

In the more general case, we contend that even when legal privacy principles do not apply for a particular use case or context, \textbf{they can still provide valuable guidelines for the design of \PPAs}.
For instance, assisting users with making informed, unambiguous, and explicit privacy decisions (as required for consent) may foster trust in \PPAs even when the privacy decision does not formally constitute consent.
Also, usable explanations of the risks and implications when using an \PPA can in general help raise awareness among users.\\

\noindent\fbox{
    \begin{minipage}{.9\linewidth}
        \setlength\parskip{1em}
        \textit{\textbf{Recommendation:} 
        We recommend a deeper consideration of legal requirements for the design of \PPAs.
        Such efforts could particularly amount to:
        1) meeting consent requirements when assisting on decisions related to consent, such as permission settings;
        2) the introduction of \PPAs assisting and enabling users to exercise their data subject rights; and, 
        3) the incorporation of usable explanations for the logic behind the AI-based proposed decisions, as well as information about the significance and the envisaged consequences of such automated processing for the data subject.
        %and an assessment of risks and consequences addressing related requirements by the GDPR for automated decision-making or the AI Act. 
        }
    \end{minipage}
}

% \vspace{-.4cm}
\subsection{Use of Varied Sources of Data, Accounting for Both Context and Personal Preferences}
\label{subsec:disc_sources}
Lastly, our study yielded that \PPAs leverage various sources of data (context, attitudinal data, behavioral data, type of data, content of data, and metadata), but not necessarily within the same solution.
However, utilizing several of these data sources can be a challenge in itself, as it requires careful curation of the datasets and adequate use of the AI models.
The benefits harnessed can be high, resulting potentially in higher prediction accuracy and, thus, in higher quality of privacy guidance and assistance.

We also acknowledge the difficulty of determining certain sources of data -- such as context --, or the problem of the sensitivity of data.
As mentioned in Section~\ref{subsubsec:context}, context is rarely defined.
It is, however, possible to draw inspiration for a rigorous definition from the seminal paper by Barth et al.~\cite{barth_privacy_2006} on the formalization in a logical framework of the concept of contextual integrity coined by Nissenbaum~\cite{nissenbaum_privacy_2004}.
As for the sensitivity of data, it is notably incumbent on context when, for instance, the same type of data (e.g., location) can be deemed non-sensitive in one context (e.g., at a workplace in the middle of the week), but sensitive in another (e.g., Sunday morning near a church, thereby disclosing potential religious beliefs).\\

\noindent\fbox{
    \begin{minipage}{.9\linewidth}
        \setlength\parskip{1em}
        \textit{\textbf{Recommendation:} Based on the relative singularity of data sources, we advocate for a plurality of data sources, encompassing context as much as personal preferences.}
    \end{minipage}
}

\subsection{Research Avenues}
\label{subsec:future}
In this final section of our Discussion, we explore prospective research paths on \PPAs, informed by the results of our study and the current social, technical, and legal landscape.

\subsubsection{The Future of \PPAs and Generative AI}
% \todo[inline]{Update accounts for the 2024 LLM paper? LEO: DONE}
While the uptake of generative AI, such as Large Language Models (LLMs), is undeniable, their application to \PPAs is not yet prevalent in the literature.
\textbf{Only one of the included studies leverages LLMs to enable automated privacy decisions.} This solution, referred to as PrivacyOracle, allows users to have a ``privacy firewall'' for filtering and managing personal data flows in the context of smart buildings~\cite{wang_privacyoracle_2024}.
Nonetheless, we can also acknowledge other research not included in this SLR, such as the work of Hamid et al.~\cite{hamid_genaipabench_2023} that provided a benchmark for evaluating Generative AI-based Privacy Assistants to simplify and make privacy policies more user-friendly.
Note that this work was not included in the SLR, as it only enhanced explanations but did not automate any privacy decisions.

We are perhaps one step away from having a new wave of LLM-powered \PPAs.
LLMs are great for summarizing and capturing insights from a large number of text (e.g., privacy policies, logs, traffic data, and system documentation).
We envision that if such insights become reliable enough, the user's privacy preferences could be automatically matched with a given system's privacy configurations, semi-automating decisions, providing allow/deny rules based on previous privacy settings for similar systems, etc.
In the work of Wang et al.~\cite{wang_privacyoracle_2024}, their PrivacyOracle was already achieving 98\% accuracy in identifying privacy-sensitive states from sensor data and 75\% accuracy in measuring the social acceptability of information flows.
However, such opportunities also raise a series of risks.

LLMs are intrinsically challenging to explain and lack transparency and interpretability. 
Furthermore, due to the risk that they respond with false or misleading information presented as facts (or ``hallucinate''), \textbf{they conflict directly with compliance requirements such as the GDPR data accuracy principle,\footnote{Art 5 (i) (e) GDPR indeed stipulates that data needs to be ``accurate and, where necessary, kept up to date; every reasonable step must be taken to ensure that personal data that are inaccurate, having regard to the purposes for which they are processed, are erased or rectified without delay (‘accuracy’)''.} and should therefore be incorporated into \PPAs with caution}.
Still, future research should address opportunities and challenges of designing and using LLM-based \PPAs, as well as technical and legal requirements for involving LLMs in assisting users with privacy decisions.

\subsubsection{Designing Genuinely Privacy-friendly \PPAs}
A promising yet critical avenue for future research remains \textbf{to design a genuinely privacy-friendly \PPA, with the right amount of notice} to empower users and avoid the so-called ``consent fatigue.''
This right amount of notice can be a difficult balance to achieve -- some users favor more notice than others -- but it is a crucial step for the uptake of such assistants.

The design should naturally be informed by the latest results in the academic literature~\cite{feng_design_2021}; it should carefully consider the number of notices, their content, their timing, etc.
However, it should also be complemented by usability studies conducted in the early stages of the prototype, as iterations over the design of the assistant are likely to be required to fine-tune it.

\subsubsection{Trust in the AI assistants and automation bias}
Individuals tend to overly trust AI systems and favor AI-based decision-making while ignoring contradictory information made without automation, a phenomenon known as automation bias~\cite{cummings2017automation}, which is a problem also related to the Elisa effect first described by Weizenbaum~\cite{weizenbaum1976computer}.

If the user's decisions are biased towards following a privacy recommendation proposed by a \PPA, the users' autonomy may be negatively impacted in practice. Hence, future research should examine if users may too easily trust and rely on proposed or nudged decisions by \PPAs without critically judging or adapting proposed decisions and how such a problem could be addressed by Human-Computer Interaction research.

\section{Threats to Validity}
\label{sec:validity-threats}
% \todo[inline]{Provide additional details on the selection process → @leo (essentially rolling back a longer version of the threat to validity)}

\subsection{Threat I -- Planning Limitations of the SLR}
The first threat relates to the planning of the SLR in terms of identifying the need and justification for this study. Here, we were concerned with identifying existing reviews (systematic and non-systematic) on the topic of \PPAs. The initial searches did not reveal any review studies on the topic, as described in Section~\ref{sec:slr-planning}, pointing to a significant gap in secondary research AI PPAs. The planning phase of the SLR is also critical to outline the research questions and provide the basis for an objective investigation of the studies that are being reviewed. If the RQs are not explicitly stated or omit the key topics, the literature review results can be flawed, overlooking the key information. To mitigate this threat, we outlined two RQs and objectives (Section~\ref{sec:slr-planning}). In summary, we seek to minimize any bias or limitations during the planning phase when defining the scope and objectives of this SLR. As a last step in the planning phase, the team finalized and cross-checked the study protocol to minimize the limitations of the SLR plan before proceeding to the subsequent phases.

\subsection{Threat II -- Validity of the Search Process}
Identifying and selecting the studies reviewed in the SLR are also significant processes to be observed. Selecting studies is a critical step; if any relevant papers are missed, the results of the SLR may be flawed. Therefore, we followed a stepwise process (Section~\ref{sec:slr-conducting}), starting with a literature screening and followed by a complete reading of papers. This selection process was carried out independently by two reviewers. We also performed forward and backward snowballing, looking for references to other potentially relevant studies. Also, this SLR restricts the selection of publications to four scientific databases: Scopus, Web of Science, IEEE Xplore, and ACM Digital Library. These databases were used due to their high relevancy to computer science, privacy, and data protection, as well as to maintain a feasible search space. This search process gives us confidence that we minimized limitations related to (i) excluding or overlooking relevant studies or (ii) including irrelevant studies that could impact the results and their reporting in the SLR.

\subsection{Threat III -- Potential Bias in the Synthesis Process}
Some threats should also be considered regarding the potential bias in synthesizing the data from the review and documenting the results. This means that if there are some limitations in the data synthesis, they directly impact the results of this SLR. Typical examples of such limitations could be a flawed research taxonomy and a mismatch of potential research gaps. To minimize the bias in synthesizing and reporting the results, we have created a data extraction form that uses well-known classification schemes, such as the ones proposed by Wieringa et al.~\cite{wieringa_requirements_2006} and Creswell and Creswell~\cite{creswell_research_2017}, or Arrieta et al.~\cite{arrieta_explainable_2019} for the classification of AI. Three researchers independently reviewed this data extraction form while revising the research protocol. While one of the researchers led the data extraction step, two other authors helped by cross-checking the work throughout the process for consistency. Three authors were involved in the creation of the classification scheme derived from the literature (i.e., shown in Section~\ref{fig:classification}), actively working on reviewing the list of categories for consistency through a series of meetings. Furthermore, this SLR also offers a complete replication package, enabling researchers to reproduce or extend this review (\url{https://github.com/Victor-Morel/SLR_AI_PPA}).

\section{Conclusion}
\label{sec:conclusion}
With the SLR presented in this article, we provide a classification and common vocabulary to compare and discuss \PPAs.
Although many papers (41 in our selection) have already been published on \PPAs in the last decade, they do not yet form a coherent body of literature.
\PPAs can be improved by performing standard evaluations (including their usability), integrating privacy by design in their design process, providing additional explanations for their decisions, and considering a broader range of system context and larger variety of data sources.
We hope this survey and its classification allow users and developers of \PPAs to compare different solutions and understand their pros and cons. Moreover, the recommendations should help improve \PPAs in different ways, addressing the challenges raised by AI's latest developments (including LLMs), data collection, and modern regulations.

\bibliographystyle{ieeetr}
\bibliography{APD_survey,APD_suppl}

\begin{thebibliography}{10}

\bibitem{noauthor_state_2024}
``State of {IoT} 2024: Number of connected {IoT} devices growing 13\% to 18.8
  billion globally.''

\bibitem{european_parliament_general_2016}
{European Commission}, ``{Regulation ({EU}) 2016/679 of the {European}
  {Parliament} and of the {Council} of 27 April 2016 on the protection of
  natural persons with regard to the processing of personal data and on the
  free movement of such data, and repealing {Directive} 95/46/EC ({General Data
  Protection Regulation})},'' {\em Official Journal of the European Union},
  no.~April, 2016.

\bibitem{choi_role_2018}
H.~Choi, J.~Park, and Y.~Jung, ``The role of privacy fatigue in online privacy
  behavior,'' {\em Computers in Human Behavior}, vol.~81.

\bibitem{sadeh2021personalized}
N.~Sadeh, B.~Liu, A.~Das, M.~Degeling, and F.~Schaub, ``Personalized privacy
  assistant,'' Mar.~23 2021.
\newblock US Patent 10,956,586.

\bibitem{eu_ai_act_2024}
{European Commission}, ``Regulation ({EU}) 2024/1689 of the {European}
  {Parliament} and of the {Council} of 13 june 2024 laying down harmonised
  rules on artificial intelligence and amending regulations ({EC}) no 300/2008,
  ({EU}) no 167/2013, ({EU}) no 168/2013, ({EU}) 2018/858, ({EU}) 2018/1139 and
  ({EU}) 2019/2144 and {Directives 2014/90/EU}, ({EU}) 2016/797 and ({EU})
  2020/1828 ({Artificial Intelligence Act}) (text with {EEA} relevance),'' {\em
  Official Journal of the European Union}, no.~June, 2024.

\bibitem{westin_privacy_1968}
A.~F. Westin, ``Privacy and freedom,'' {\em Washington and Lee Law Review},
  vol.~25, no.~1.

\bibitem{Art29WP13}
{Art. 29 Working Party}, ``{Opinion 2/2013: Apps on smart devices},'' 2013.

\bibitem{human_data_2022}
S.~Human, H.~J. Pandit, V.~Morel, C.~Santos, M.~Degeling, A.~Rossi, W.~Botes,
  V.~Jesus, and I.~Kamara, ``Data protection and consenting communication
  mechanisms: Current open proposals and challenges,'' in {\em 2022 {IEEE}
  European Symposium on Security and Privacy Workshops ({EuroS}\&{PW})},
  {IEEE}.

\bibitem{CCPA_2018}
``California consumer privacy act of 2018.''
  \url{https://oag.ca.gov/privacy/ccpa}, 2018.
\newblock Cal. Civ. Code § 1798.100 et seq.

\bibitem{cranor_platform_2002}
L.~Cranor, M.~Langheinrich, M.~Marchiori, M.~Presler-Marshall, and J.~Reagle,
  ``The platform for privacy preferences 1.0 ({P3P1}. 0) specification,'' {\em
  W3C recommendation}, vol.~16, 2002.

\bibitem{DNT_W3C_2019}
W.~W. W.~C. (W3C), ``Tracking preference expression (dnt).''
  \url{https://www.w3.org/TR/tracking-dnt/}, 2019.
\newblock W3C Recommendation.

\bibitem{russell_artificial_2016}
S.~J. Russell and P.~Norvig, {\em Artificial intelligence: a modern approach}.
\newblock Pearson.

\bibitem{arrieta_explainable_2019}
A.~B. Arrieta, N.~D{\'\i}az-Rodr{\'\i}guez, J.~Del~Ser, A.~Bennetot, S.~Tabik,
  A.~Barbado, S.~Garc{\'\i}a, S.~Gil-L{\'o}pez, D.~Molina, R.~Benjamins, {\em
  et~al.}, ``Explainable artificial intelligence ({XAI}): Concepts, taxonomies,
  opportunities and challenges toward responsible {AI},'' 2020.

\bibitem{panigutti_role_2023}
C.~Panigutti, R.~Hamon, I.~Hupont, D.~Fernandez~Llorca, D.~Fano~Yela,
  H.~Junklewitz, S.~Scalzo, G.~Mazzini, I.~Sanchez, J.~Soler~Garrido, {\em
  et~al.}, ``The role of explainable {AI} in the context of the {AI} {A}ct,''
  pp.~1139--1150, 2023.

\bibitem{puiutta_explainable_2020}
E.~Puiutta and E.~M. Veith, ``Explainable reinforcement learning: A survey,''
  2020.

\bibitem{kitchenham2004procedures}
B.~Kitchenham, ``Procedures for performing systematic reviews,'' Tech. Rep.
  2004, Keele, UK, Keele University, 2004.

\bibitem{marky_decide_2024}
K.~Marky, A.~Stöver, S.~Prange, K.~Bleck, P.~Gerber, V.~Zimmermann,
  F.~Müller, F.~Alt, and M.~Mühlhäuser, ``Decide yourself or delegate - user
  preferences regarding the autonomy of personal privacy assistants in private
  {IoT}-equipped environments,'' in {\em Proceedings of the {CHI} Conference on
  Human Factors in Computing Systems}, pp.~1--20, {ACM}.

\bibitem{bollinger_automating_2022}
D.~Bollinger, K.~Kubicek, C.~Cotrini, and D.~Basin, ``Automating cookie consent
  and {GDPR} violation detection,'' {\em 31st USENIX Security Symposium (USENIX
  Security 22)}, pp.~2893--2910, 2022.

\bibitem{wohlin_guidelines_2014}
C.~Wohlin, ``Guidelines for snowballing in systematic literature studies and a
  replication in software engineering,'' in {\em Proceedings of the 18th
  International Conference on Evaluation and Assessment in Software
  Engineering}, {ACM}.

\bibitem{kuhrmann_software_2016}
M.~Kuhrmann, P.~Diebold, and J.~Münch, ``Software process improvement: a
  systematic mapping study on the state of the art,'' {\em {PeerJ} Computer
  Science}, vol.~2.

\bibitem{shaw_writing_2003}
M.~Shaw, ``Writing good software engineering research papers,'' in {\em 25th
  International Conference on Software Engineering, 2003. Proceedings.},
  {IEEE}.

\bibitem{xie_location_2014}
J.~Xie, B.~P. Knijnenburg, and H.~Jin, ``Location sharing privacy preference:
  analysis and personalized recommendation,'' in {\em Proceedings of the 19th
  international conference on Intelligent User Interfaces}, {ACM}.

\bibitem{apolinarski_automating_2015}
W.~Apolinarski, M.~Handte, and P.~J. Marron, ``Automating the generation of
  privacy policies for context-sharing applications,'' in {\em 2015
  International Conference on Intelligent Environments}, {IEEE}.

\bibitem{hirschprung_simplifying_2015}
R.~Hirschprung, E.~Toch, and O.~Maimon, ``Simplifying data disclosure
  configurations in a cloud computing environment,'' {\em {ACM} Transactions on
  Intelligent Systems and Technology}, vol.~6, no.~3.

\bibitem{squicciarini_privacy_2015}
A.~C. Squicciarini, D.~Lin, S.~Sundareswaran, and J.~Wede, ``Privacy policy
  inference of user-uploaded images on content sharing sites,'' {\em {IEEE}
  Transactions on Knowledge and Data Engineering}, vol.~27, no.~1.

\bibitem{liu_follow_2016}
B.~Liu, M.~S. Andersen, F.~Schaub, H.~Almuhimedi, S.~A. Zhang, N.~Sadeh,
  Y.~Agarwal, and A.~Acquisti, ``Follow my recommendations: A personalized
  privacy assistant for mobile app permissions,'' in {\em Twelfth symposium on
  usable privacy and security (SOUPS 2016)}, pp.~27--41, 2016.

\bibitem{albertini_privacy_2016}
D.~A. Albertini, B.~Carminati, and E.~Ferrari, ``Privacy settings recommender
  for online social network,'' in {\em 2016 {IEEE} 2nd International Conference
  on Collaboration and Internet Computing ({CIC})}, {IEEE}.

\bibitem{dong_ppm_2016}
C.~Dong, H.~Jin, and B.~P. Knijnenburg, ``{PPM}: {A} privacy prediction model
  for online social networks,'' 2016.

\bibitem{baarslag_automated_2017}
T.~Baarslag, A.~T. Alan, R.~Gomer, M.~Alam, C.~Perera, E.~H. Gerding, and
  m.~schraefel, ``An automated negotiation agent for permission management,''
  pp.~380--390, 2017.

\bibitem{fogues_sosharp_2017}
R.~L. Fogues, P.~K. Murukannaiah, J.~M. Such, and M.~P. Singh, ``So{S}harp:
  Recommending sharing policies in multiuser privacy scenarios,'' {\em IEEE
  Internet Computing}, vol.~21, no.~6, pp.~28--36, 2017.

\bibitem{zhong_group-based_2017}
H.~Zhong, A.~Squicciarini, D.~Miller, and C.~Caragea, ``A group-based
  personalized model for image privacy classification and labeling,'' in {\em
  Proceedings of the Twenty-Sixth International Joint Conference on Artificial
  Intelligence}, International Joint Conferences on Artificial Intelligence
  Organization.

\bibitem{misra_pacman_2017}
G.~Misra and J.~M. Such, ``{PACMAN}: Personal agent for access control in
  social media,'' {\em {IEEE} Internet Computing}, vol.~21, no.~6.

\bibitem{camp_easing_2017}
T.~Nakamura, S.~Kiyomoto, W.~B. Tesfay, and J.~Serna, ``Easing the burden of
  setting privacy preferences: a machine learning approach,'' in {\em
  Information Systems Security and Privacy} (O.~Camp, S.~Furnell, and P.~Mori,
  eds.), vol.~691, Springer International Publishing.
\newblock Series Title: Communications in Computer and Information Science.

\bibitem{olejnik_smarper_2017}
K.~Olejnik, I.~Dacosta, J.~S. Machado, K.~Huguenin, M.~E. Khan, and J.-P.
  Hubaux, ``{SmarPer}: Context-aware and automatic runtime-permissions for
  mobile devices,'' in {\em 2017 {IEEE} Symposium on Security and Privacy
  ({SP})}, {IEEE}.

\bibitem{das_personalized_2018}
A.~Das, M.~Degeling, D.~Smullen, and N.~Sadeh, ``Personalized privacy
  assistants for the internet of things: Providing users with notice and
  choice,'' {\em {IEEE} Pervasive Computing}, vol.~17, no.~3.

\bibitem{tan_context-perceptual_2018}
H.-Z. Tan, W.~Zhao, and H.-H. Shen, ``A context-perceptual privacy protection
  approach on android devices,'' in {\em 2018 {IEEE} International Conference
  on Communications ({ICC})}, {IEEE}.

\bibitem{wijesekera_contextualizing_2018}
P.~Wijesekera, J.~Reardon, I.~Reyes, L.~Tsai, J.-W. Chen, N.~Good, D.~Wagner,
  K.~Beznosov, and S.~Egelman, ``Contextualizing privacy decisions for better
  prediction (and protection),'' in {\em Proceedings of the 2018 {CHI}
  Conference on Human Factors in Computing Systems}, {ACM}.

\bibitem{yu_leveraging_2018}
J.~Yu, Z.~Kuang, B.~Zhang, W.~Zhang, D.~Lin, and J.~Fan, ``Leveraging content
  sensitiveness and user trustworthiness to recommend fine-grained privacy
  settings for social image sharing,'' {\em {IEEE} Transactions on Information
  Forensics and Security}, vol.~13, no.~5.

\bibitem{bahirat_data-driven_2018}
P.~Bahirat, Y.~He, A.~Menon, and B.~Knijnenburg, ``A data-driven approach to
  developing {IoT} privacy-setting interfaces,'' in {\em 23rd International
  Conference on Intelligent User Interfaces}, {ACM}.

\bibitem{raber_retailio_2018}
F.~Raber, D.~Ziemann, and A.~Krueger, ``The 'retailio' privacy wizard:
  Assisting users with privacy settings for intelligent retail stores,'' in
  {\em Proceedings 3rd European Workshop on Usable Security}, Internet Society.

\bibitem{klingensmith_hypervisor-based_2019}
N.~Klingensmith, Y.~Kim, and S.~Banerjee, ``A hypervisor-based privacy agent
  for mobile and {IoT} systems,'' in {\em Proceedings of the 20th International
  Workshop on Mobile Computing Systems and Applications}, {ACM}.

\bibitem{barbosa_what_2019}
N.~M. Barbosa, J.~S. Park, Y.~Yao, and Y.~Wang, ``\textit{“What if?”}
  predicting individual users’ smart home privacy preferences and their
  changes,'' {\em Proceedings on Privacy Enhancing Technologies}, vol.~2019,
  no.~4.

\bibitem{alom_helping_2019}
M.~Z. Alom, B.~Carminati, and E.~Ferrari, ``Helping users managing
  context-based privacy preferences,'' in {\em 2019 {IEEE} International
  Conference on Services Computing ({SCC})}, {IEEE}.

\bibitem{alom_adapting_2019}
M.~Z. Alom, B.~Carminati, and E.~Ferrari, ``Adapting users' privacy preferences
  in smart environments,'' in {\em 2019 {IEEE} International Congress on
  Internet of Things ({ICIOT})}, {IEEE}.

\bibitem{barolli_selflearning_2020}
M.~Kasaraneni and J.~P. Thomas, ``A self–learning personal privacy
  assistant,'' in {\em Advanced Information Networking and Applications}
  (L.~Barolli, F.~Amato, F.~Moscato, T.~Enokido, and M.~Takizawa, eds.),
  vol.~1151, Springer International Publishing.
\newblock Series Title: Advances in Intelligent Systems and Computing.

\bibitem{kaur_smart_2020}
H.~Kaur, I.~Echizen, and R.~Kumar, ``Smart data agent for preserving location
  privacy,'' in {\em 2020 {IEEE} Symposium Series on Computational Intelligence
  ({SSCI})}, {IEEE}.

\bibitem{herrero_automatic_2021}
V.~Botti-Cebri{\'a}, E.~Del~Val, and A.~Garc{\'\i}a-Fornes, ``Automatic
  detection of sensitive information in educative social networks,'' in {\em
  13th International Conference on Computational Intelligence in Security for
  Information Systems (CISIS 2020) 12}, pp.~184--194, 2021.

\bibitem{kokciyan_turp_2020}
N.~Kokciyan and P.~Yolum, ``{TURP}: Managing trust for regulating privacy in
  internet of things,'' {\em {IEEE} Internet Computing}, vol.~24, no.~6.

\bibitem{sanchez_recommendation_2020}
O.~R. Sanchez, I.~Torre, Y.~He, and B.~P. Knijnenburg, ``A recommendation
  approach for user privacy preferences in the fitness domain,'' {\em User
  Modeling and User-Adapted Interaction}, vol.~30, no.~3.

\bibitem{barolli_reinforcement_2021}
H.~Kaur, R.~Kumar, and I.~Echizen, ``Reinforcement learning based smart data
  agent for location privacy,'' in {\em International Conference on Advanced
  Information Networking and Applications}, pp.~657--671, 2021.

\bibitem{lobner_explainable_2021}
S.~Lobner, W.~B. Tesfay, T.~Nakamura, and S.~Pape, ``Explainable machine
  learning for default privacy setting prediction,'' {\em {IEEE} Access},
  vol.~9.

\bibitem{filipczuk_automated_2022}
D.~Filipczuk, T.~Baarslag, E.~H. Gerding, and M.~C. Schraefel, ``Automated
  privacy negotiations with preference uncertainty,'' {\em Autonomous Agents
  and Multi-Agent Systems}, vol.~36, no.~2.

\bibitem{hirschprung_game_2022}
R.~S. Hirschprung and S.~Alkoby, ``A game theory approach for assisting humans
  in online information-sharing,'' {\em Information}, vol.~13, no.~4.

\bibitem{kokciyan_taking_2022}
N.~K{\"o}kciyan, P.~Yolum, {\em et~al.}, ``Taking situation-based privacy
  decisions: Privacy assistants working with humans.,'' in {\em IJCAI},
  pp.~703--709, 2022.

\bibitem{ulusoy_panola_2022}
O.~Ulusoy and P.~Yolum, ``{PANOLA}: A personal assistant for supporting users
  in preserving privacy,'' {\em {ACM} Transactions on Internet Technology},
  vol.~22, no.~1.

\bibitem{zhan_model_2022}
X.~Zhan, S.~Sarkadi, N.~Criado, and J.~Such, ``A model for governing
  information sharing in smart assistants,'' in {\em Proceedings of the 2022
  {AAAI}/{ACM} Conference on {AI}, Ethics, and Society}, {ACM}.

\bibitem{brandao_prediction_2022}
A.~Brandão, R.~Mendes, and J.~P. Vilela, ``Prediction of mobile app privacy
  preferences with user profiles via federated learning,'' in {\em Proceedings
  of the Twelfth {ACM} Conference on Data and Application Security and
  Privacy}, {ACM}.

\bibitem{mendes_enhancing_2022}
R.~Mendes, M.~Cunha, J.~P. Vilela, and A.~R. Beresford, ``Enhancing user
  privacy in mobile devices through prediction of privacy preferences,'' in
  {\em European Symposium on Research in Computer Security}, pp.~153--172,
  Springer, 2022.

\bibitem{shanmugarasa_automated_2022}
Y.~Shanmugarasa, H.-y. Paik, S.~S. Kanhere, and L.~Zhu, ``Automated privacy
  preferences for smart home data sharing using personal data stores,'' {\em
  {IEEE} Security \& Privacy}, vol.~20, no.~1.

\bibitem{ayci_uncertainty-aware_2023}
G.~Ayci, M.~Sensoy, A.~Özgür, and P.~Yolum, ``Uncertainty-aware personal
  assistant for making personalized privacy decisions,'' {\em {ACM}
  Transactions on Internet Technology}, vol.~23, no.~1.

\bibitem{serramia_predicting_2023}
M.~S. Amoros, W.~Seymour, N.~Criado, and M.~Luck, ``Predicting privacy
  preferences for smart devices as norms,'' in {\em The 22nd International
  Conference on Autonomous Agents and Multiagent Systems}, International
  Foundation for Autonomous Agents and Multiagent Systems (IFAAMAS), 2023.

\bibitem{wang_privacyoracle_2024}
B.~Wang, L.~A. Garcia, and M.~Srivastava, ``{PrivacyOracle}: Configuring sensor
  privacy firewalls with large language models in smart built environments,''
  in {\em 2024 {IEEE} Security and Privacy Workshops ({SPW})}, pp.~239--245,
  {IEEE}.

\bibitem{cebma_cat_2025}
{CEBMa}, ``{CAT} manager app.''

\bibitem{wieringa_requirements_2006}
R.~Wieringa, N.~Maiden, N.~Mead, and C.~Rolland, ``Requirements engineering
  paper classification and evaluation criteria: a proposal and a discussion,''
  {\em Requirements Engineering}, vol.~11, no.~1.

\bibitem{nissenbaum_privacy_2004}
H.~Nissenbaum, ``Privacy as contextual integrity,'' {\em Wash. L. Rev.},
  vol.~79, p.~119, 2004.

\bibitem{shokri2017membership}
R.~Shokri, M.~Stronati, C.~Song, and V.~Shmatikov, ``Membership inference
  attacks against machine learning models,'' in {\em 2017 IEEE symposium on
  security and privacy (SP)}, pp.~3--18, IEEE, 2017.

\bibitem{pallas_privacy_2024}
F.~Pallas, K.~Koerner, I.~Barberá, J.-H. Hoepman, M.~Jensen, N.~R. Narla,
  N.~Samarin, M.-R. Ulbricht, I.~Wagner, K.~Wuyts, and C.~Zimmermann, ``Privacy
  engineering from principles to practice: A roadmap,'' {\em {IEEE} Security \&
  Privacy}, vol.~22, no.~2.

\bibitem{schaub_design_2015}
F.~Schaub, R.~Balebako, A.~L. Durity, and L.~F. Cranor, ``A design space for
  effective privacy notices,'' in {\em Eleventh Symposium On Usable Privacy and
  Security ({SOUPS} 2015)}.

\bibitem{morel_automating_2023}
V.~Morel and S.~Fischer-Hübner, ``Automating privacy decisions-where to draw
  the line?,'' in {\em 2023 {IEEE} European Symposium on Security and Privacy
  Workshops ({EuroS}\&{PW})}, {IEEE}.

\bibitem{mothukuri2021survey}
V.~Mothukuri, R.~M. Parizi, S.~Pouriyeh, Y.~Huang, A.~Dehghantanha, and
  G.~Srivastava, ``A survey on security and privacy of federated learning,''
  {\em Future Generation Computer Systems}, vol.~115, pp.~619--640, 2021.

\bibitem{jebreel2022enhanced}
N.~M. Jebreel, J.~Domingo-Ferrer, A.~Blanco-Justicia, and D.~S{\'a}nchez,
  ``Enhanced security and privacy via fragmented federated learning,'' {\em
  IEEE transactions on neural networks and learning systems}, 2022.

\bibitem{PAPAYA2019}
B.~Bozdemir, O.~Ermis, M.~{\"O}nen, M.~Barham, B.~Rozenberg, R.~Shmelkin,
  M.~Azraoui, S.~Canard, and B.~Vialla, ``Papaya {P}roject - {D}3.1 --
  {P}reliminary {D}esign of {P}rivacy {P}reserving {D}ata {A}nalytics,'' 2019.

\bibitem{narayanan_ai_2024}
A.~Narayanan and S.~Kapoor, {\em {AI} Snake oil: what artificial intelligence
  can do, what it can’t, and how to tell the difference}.
\newblock Princeton University Press.

\bibitem{crane2006trustguide}
S.~Crane, H.~Lacoh{\'e}e, and S.~Zaba, ``Trustguide—trust in {ICT},'' {\em BT
  technology journal}, vol.~24, no.~4, pp.~69--80, 2006.

\bibitem{ifttt_ifttt_nodate}
{IFTTT}, ``{IFTTT} - automate business \& home.''

\bibitem{barth_privacy_2006}
A.~Barth, A.~Datta, J.~C. Mitchell, and H.~Nissenbaum, ``Privacy and contextual
  integrity: Framework and applications,'' in {\em 2006 {IEEE} symposium on
  security and privacy (S\&P'06)}, {IEEE}.

\bibitem{hamid_genaipabench_2023}
A.~Hamid, H.~R. Samidi, P.~Pappachan, T.~Finin, and R.~Yus, ``Gen{AIPAB}ench: A
  {B}enchmark for {G}enerative {AI}-based {P}rivacy {A}ssistants,'' 2024.

\bibitem{feng_design_2021}
Y.~Feng, Y.~Yao, and N.~Sadeh, ``A design space for privacy choices: {T}owards
  meaningful privacy control in the {I}nternet of {T}hings,'' in {\em
  Proceedings of the 2021 {CHI} Conference on Human Factors in Computing
  Systems}.

\bibitem{cummings2017automation}
M.~L. Cummings, ``Automation bias in intelligent time critical decision support
  systems,'' in {\em Decision making in aviation}, pp.~289--294, Routledge,
  2017.

\bibitem{weizenbaum1976computer}
J.~Weizenbaum, ``Computer power and human reason: From judgment to
  calculation,'' 1976.

\bibitem{creswell_research_2017}
J.~W. Creswell and J.~D. Creswell, {\em Research design: Qualitative,
  quantitative, and mixed methods approaches}.
\newblock United States: Sage publications, 2017.

\end{thebibliography}

\newpage

\begin{IEEEbiographynophoto}{Victor Morel} holds a PhD in computer science from Inria - Privatics, and an French national engineering degree from Polytech' Lyon, France.

His research interests include privacy and data protection, networks security, usability and Human-Computer Interactions, applied cryptography, and ethics of technology in a broad manner. 
He appreciates sober, hackable, and accessible technology that works.

He is currently working as a postdoctoral researcher in the Security \& Privacy Lab at Chalmers University of Technology on usable privacy for IoT applications. 
He is also a selected member of the EDPB’s support pool of experts.
\end{IEEEbiographynophoto}

\begin{IEEEbiographynophoto}{Leonardo H. Iwaya} is an Associate Senior Lecturer in the Department of Mathematics and Computer Science at Karlstad University, Sweden. He obtained a Ph.D. degree in computer science from Karlstad University, Sweden. He also holds an M.Sc. in electrical engineering from the University of S\~{a}o Paulo and a B.Sc. in computer science from Santa Catarina State University, Brazil. From 2011 to 2014, he was a Research Assistant with the Laboratory of Computer Networks and Architecture (LARC), PCS-EPUSP. From 2019 to 2021, he worked as a Postdoctoral Researcher with the School of Computer Science at the University of Adelaide, Australia, as part of the Cyber Security Cooperative Research Centre (CSCRC). During 2021 and 2022, he worked as a Postdoctoral Researcher with the Interdisciplinary Research Group on Knowledge, Learning and Organizational Memory (KLOM) at the Federal University of Santa Catarina. He currently works with the Privacy \& Security (PriSec) Research Group at Karlstad University, contributing to projects such as DHINO and DigitalWell Arena. His research interests include privacy engineering, cybersecurity, human factors, mobile and ubiquitous health systems, and the privacy impacts of new technologies.
\end{IEEEbiographynophoto}

% \newpage

%If you do not have or do not want to include a photo, you can use IEEEbiographynophoto as shown below:

\begin{IEEEbiographynophoto}{Simone Fischer-Hübner} has been a Full Professor at Karlstad University since June 2000, where is the head of the Privacy \& Security (PriSec) research group. Moreover, since April 2022 she is a part-time Guest Professor at Chalmers University of Technology.

She has been conducting research in privacy, cyber security and privacy-enhancing technologies for more than 35 years.

She is the Swedish representative of IFIP (International Federation for Information Processing) TC 11 (Technical Committee on Information Security and Privacy Protection) and is the IFIP TC 11 vice chair.

Moreover, she is a member of the board for the Privacy Enhancing Technology Symposia (PETS) and for the NordSec conferences, and coordinator of the Swedish IT Security Network for PhD Students (SWITS) and of the Swedish Industrial Graduate School for Cybersecurity (SIGS-CyberSec).

She is a senior IEEE member and received the IFIP Silver Core Award in 2001, the IFIP William Winsborough Award in 2016 and an honorary doctorate from the Chalmers University of Technology in 2020.

\end{IEEEbiographynophoto}

\EOD

\end{document}